\documentclass[aps, reprint, longbibliography]{revtex4-2}
\usepackage{graphicx, natbib, amsmath}
\usepackage{xcolor} 
\usepackage{subfigure, amssymb}
\usepackage{float} 
\usepackage{color}

\DeclareGraphicsExtensions{.pdf,.eps,.png,.jpeg,.mps}
\graphicspath{ {Figures/} }

\begin{document}

\title{Spontaneous spiral patterns etched on Germanium }
\author{Yilin Wong}
\author{Giovanni Zocchi}
\email{zocchi@physics.ucla.edu}
\affiliation{Department of Physics and Astronomy, University of California - Los Angeles}

\begin{abstract}
\noindent Thin metal film on Germanium, in the presence of water, results in a remarkable pattern forming system. 
Here we present an analysis of spirals spontaneously etched on the Ge surface. 
We obtain measurements of the growth dynamics of the spirals and mesurements of the local strain field in the metal film. 
Both indicate that the near geometric order of the pattern originates from the unique far field 
of a singularity - a crystal defect. The measured engraving profile is found in quantitative agreement with 
a model of metal catalyzed corrosion of the Ge surface. Specifically, local etch depth is inversely proportional to the normal 
velocity of the Ge-metal contact line. The growth mechanism combines crack propagation, 
reaction diffusion dynamics, and thin film mechanical instabilities, and illustrates how a defect's long range field 
can impose geometric order in a nonequilibrium growth process. General features relevant 
to other pattern forming systems are the coupling of chemistry and mechanics and the singularity driven order.  
\end{abstract}


\maketitle


\section{Introduction} 

\begin{figure}
	\centering
	\includegraphics[width=2.8in]{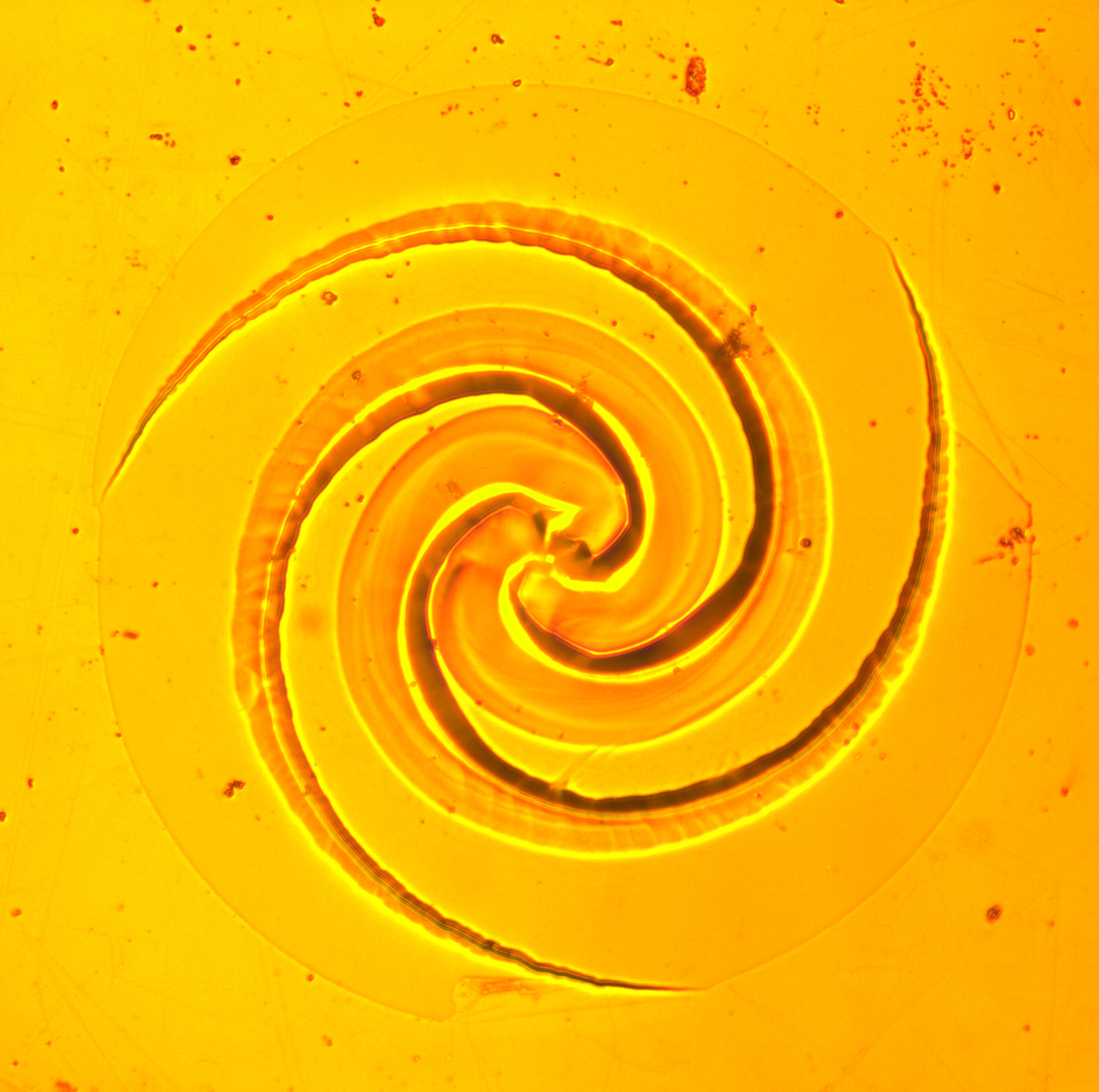}
	\caption{A three-arms Logarithmic spiral etched on the surface of Ge by the spontaneous self-organized process described 
	in the text. The overall diameter of the structure is $680 \, \mu m$. In this optical microscopy picture 
	of the dry sample, taken in reflected light, flat areas appear brighter and sloping areas darker. The arms of the spiral 
	correspond to V-shaped grooves, the terraces in between sloping upwards as one moves away from the center, turning 
	clockwise. The tiny bright dot at the center is the floor of the central etch pit. The metal film has peeled off from the structure, 
	exposing a circular disk of bare Ge surface; outside the contour of the disk the metal film still covers the surface. The metal film 
	thicknes for this sample was $3 \, nm$ Cr and $4 \, nm$ Au.}
	\label{fig:Log_spiral}
\end{figure}

\begin{figure}[H]
	\centering
	\includegraphics[width=2.8in]{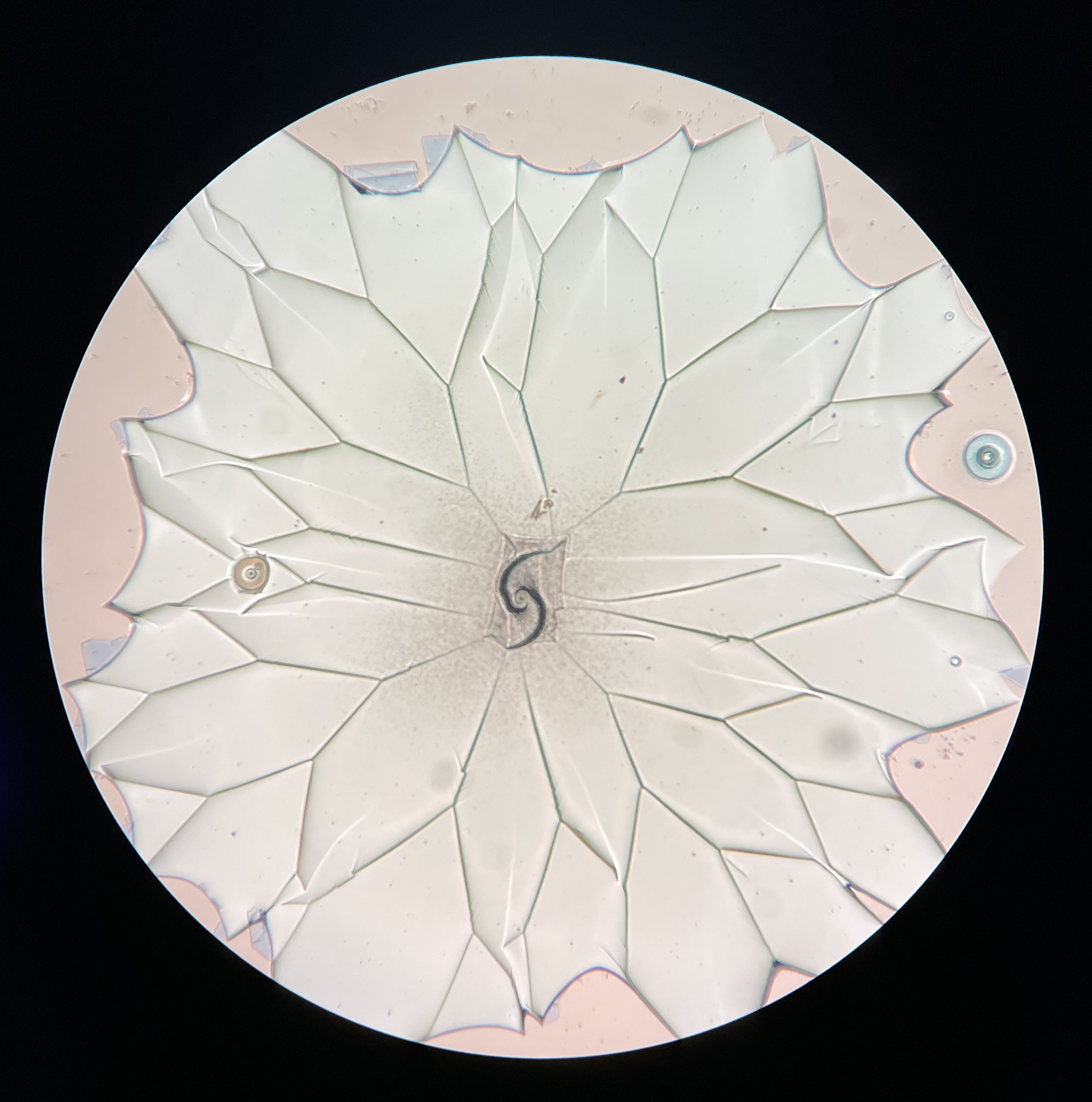}
	\caption{A lotus flower pattern produced by the same system as Fig. \ref{fig:Log_spiral}, with slight changes in parameters 
	(metal film thickness $20 \, nm$ Cr and $4 \, nm$ Au ). Reflection microscopy of the dried sample. 
	The diameter of the circular disk is $3 \, mm$. The metal film has peeled off from the pattern, 
	whereas it is still covering the Ge surface elsewhere (pinkish regions at the periphery). The ``petals'' of the 
	lotus flower are dome-shaped, with the dark lines corresponding to deeper grooves. The pattern emanates from a central 
	etch pit; it starts as a spiral and further away from the center transitions to the lotus flower. }
	\label{fig:Lotus_flower}
\end{figure}

\noindent Aesthetic sense abhors randomness and values intrinsic order. Science, concerned 
with the hidden order in the world, also values beauty: we display this connection   
in Fig. \ref{fig:Log_spiral}. This three arms spiral pattern, $\sim 680 \, \mu m$ in size, is etched on the surface of a Ge chip 
through a spontaneous process which is the subject of this paper. The system giving rise to such patterns is quite simple; 
we described it in a first report recently \cite{Yilin_3}. It consists of a Ge wafer with (100) surface orientation on which 
a thin metal film has been evaporated: a Cr layer of order $\sim 10 \, nm$ thick followed by $\sim 4 \, nm$ of Au 
(see Mat. $\&$ Met. for details). A drop of mild etching solution (water with phosphate buffer $1 \, M$ at pH 4) is deposited 
on the metallized surface, and the system is incubated for 24 - 48 hrs at room temperature. During this time, hundreds of 
patterns such as the spiral shown in Fig. \ref{fig:Log_spiral} may form at seemingly random locations on a chip a few square cm 
in size. There are several recurrent regular patterns: the Logarithmic spiral (Fig. \ref{fig:Log_spiral}), with any number of arms 
between 1 and 6, the Archimedean spiral and the radial pattern decribed previously \cite{Yilin_3}, and the ``lotus flower'' pattern 
of Fig. \ref{fig:Lotus_flower}. One chip will contain mostly or even exclusively one kind of pattern; however we have not yet been able 
to define experimental control parameters which select one specific pattern. Part of the reason will be uncovered in the following.  

\noindent The different patterns have one thing in common: they all without exception emanate from an etch pit at the center. 
The etch pit has the shape of an inverted truncated square pyramid \cite{Rhodes1957, Yilin_3} $10 - 20 \, \mu m$ 
in size and $5 - 10 \, \mu m$ deep; the sides are alligned with the crystal axes; it is believed to form at the location of a defect 
in the crystal lattice \cite{Amelinckx_Book}. The optical microscope picture of Fig. \ref{fig:Log_spiral} is obtained in reflected light 
with the dry sample. The area of the circular disk visible with faint contours is the bare Ge surface, where the metal film has 
peeled off; outside this area the metal film covers the surface. The floor of the etch pit is visible as a tiny bright dot at the center 
of the structure (for a SEM picture showing the etch pit in detail see \cite{Yilin_3}). 
Contrary to appearence, the spiral arms correspond to grooves in the Ge, not ridges, the whole structure being engraved below 
the level of the unetched surface. If we follow a spiral arm moving outwards from the center (i.e. turning 
counter clockwise), or also if we follow the widening strip between two arms, we are ``climbing'' towards the level of the unetched 
surface. Thus the deeper parts of the structure are at the center, the shallower parts at the periphery. The process that forms 
this remarkable structure depends on the fact that the etching reaction (the oxidation of Ge) is catalyzed by the metal film \cite{Yilin_3}. 
The dynamics of the chemical etching reaction is coupled to the mechanical instabilities of the metal film catalyst as material 
is removed by the etching reaction and the metal film delaminates from the Ge surface. The novelty and interest of this pattern 
forming system results from this coupling between mechanics and chemical reaction. Indeed, part of the dynamics results from 
crack propagation in the metal film and at the Ge - metal interface. 

In this paper, we focus on one type of pattern, the Logarithmic spiral. We have obtained a time lapse video recording of the growth 
of a spiral (see Supp. Mat.). From the video, we obtain the growth dynamics in quantitative detail. We also obtain an estimate 
of the strain field in the metal film. Remarkably, both measurements can be traced back to an initial singularity which seems 
to govern the formation of the pattern. We then present quantitative topographic maps of the etched surface and relate them 
to the mechanism of pattern formation. Specifically, we construct a model based on the assumption that the local amount 
of etching is proportional to the residence time of the (moving) Ge-metal contact line. We show that this model describes the 
topographic profiles fairly well. It validates the notion that catalysis occurs at the Ge-metal contact line, and aids in forming 
a consistent picture of the mechanism of pattern formation. \\ 
Beyond the specific mechanisms, are there features that carry over to other pattern forming systems ? We argue in the Discussion 
that the present system provides a remarkable example of the role of singularities in developing ordered structures in a class 
of nonequilibrium systems, which include some fluid dynamic flows and possibly structure in the Universe at the galactic scale. \\ 

Before delving into the results of this study, we will briefly mention some background.  
The discoveries of Turing \cite{Turing1952} and Belusov and Zhabotinsky \cite{Zhabotinsky1964, Zhabotinsky1991} 
ushered in pattern formation as a field of physics. 
In the second half of the last century the field developed in parallel with studies in nonlinear dynamics and dynamical systems 
theory. Model systems were drawn from materials science (e.g. crystal growth \cite{Langer1980}) 
and fluid flows \cite{Gallaire2017}, particularly thermal convection \cite{Swift1977, Cross1993}. One general approach 
on the theory side proved to be amplitude equations \cite{Cross_Book}. Inspired by biological form \cite{Darcy_Thompson_Book}, 
Alan Turing \cite{Turing1952} had pioneered the theoretical understanding of patterns associated with chemical reactions 
\cite{Winfree1972} (later termed reaction-diffusion systems). These are also associated to developments 
in nonequilibrium thermodynamics initiated by the Brussels school \cite{Prigogine1967, Prigogine1968}.  
Specifically related to the present system is the technique of Metal Assisted Chemical Etching (MACE \cite{Huang2011}). 
The method was developed to produce porous Si for optoelectronics applications \cite{Li2000}, and for 
high throughput fabrication of 3D structures on semiconductor surfaces \cite{Hildreth2009, Kawase2013, Huang2011}. 
In MACE, noble metal nanoparticles or thin films are deposited on the semiconductor surface. 
In the presence of oxidizing agents such as $HF$ or $H_2 O_2$ the metal catalyzes etching of the semiconductor, 
creating 3D patterns such as deep wells \cite{Hildreth2009}. Etching patterns such as spirals can be designed  
by shaping the metal catalyst \cite{Hildreth2012}. High aspect ratio 3D structures can be obtained because the nano particles 
``travel'' with the etching front \cite{Rykaczewski2011}. 
Also related to our system is the topic of crack propagation in thin films \cite{Yuse1993, Fineberg1991}, specifically the 
demonstration of regular crack patterns arising in substrate supported films \cite{Marthelot2014}. 
Residual tensile strain in a glassy $\mu m$ thick film drives crack propagation in the latter work, 
which is spin-on-glass on Si substrates. They observe Archimedean spirals and zig zag (``crescent alleys'') patterns. 
They construct a model in which two nearby cracks interact through the delamination area in between. The result is 
a set of equations which describe the tendency of cracks to propagate parallel to each other, forming spirals or crescent alleys, 
depending on initial conditions. The related topic of thin film mechanical instabilities \cite{Bowden1998} also intersects the present work.

\section {Results} 
\begin{figure}[H]
	\centering
	\includegraphics[width=2.0 in]{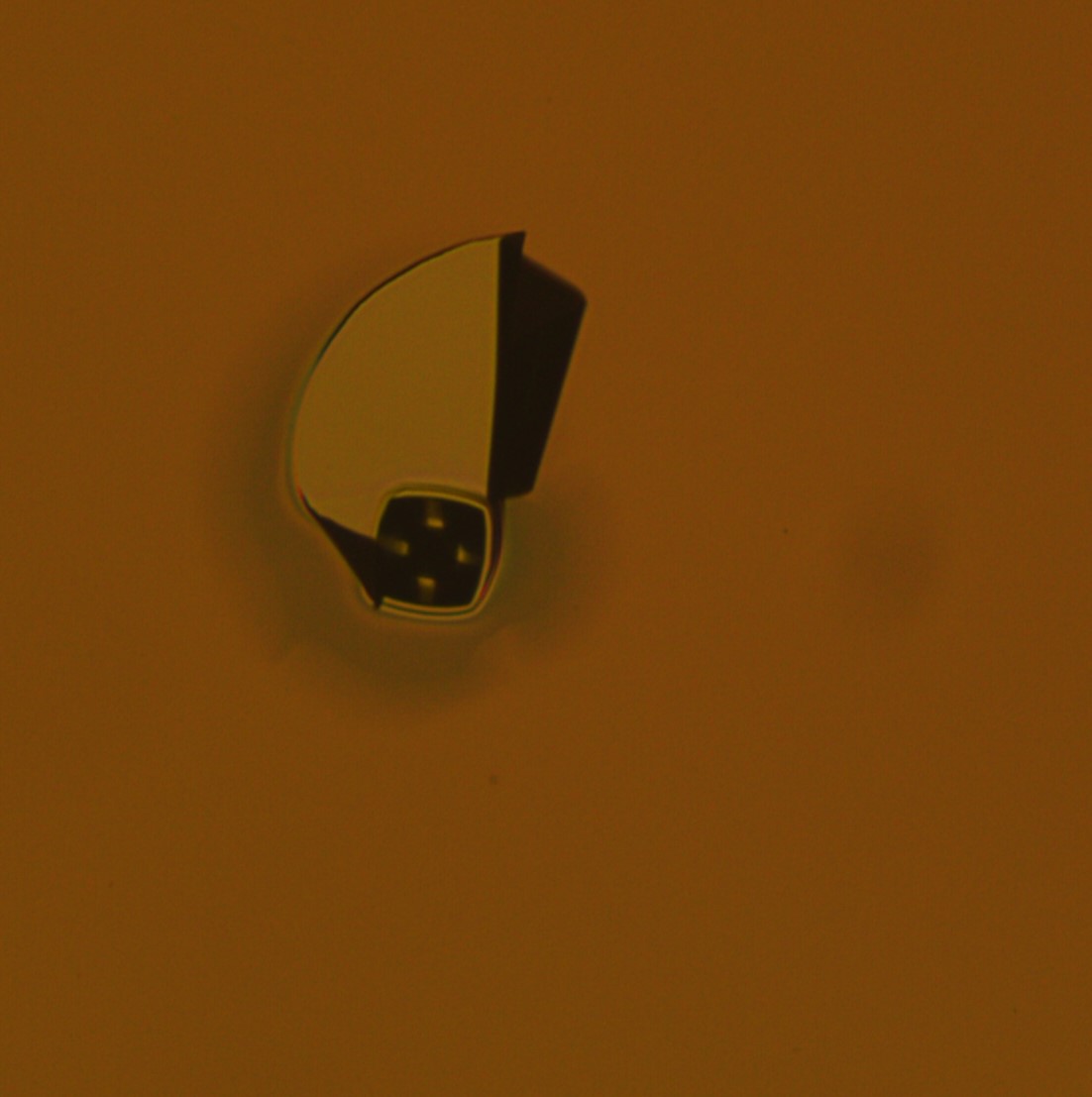}
	\includegraphics[width=2.0 in]{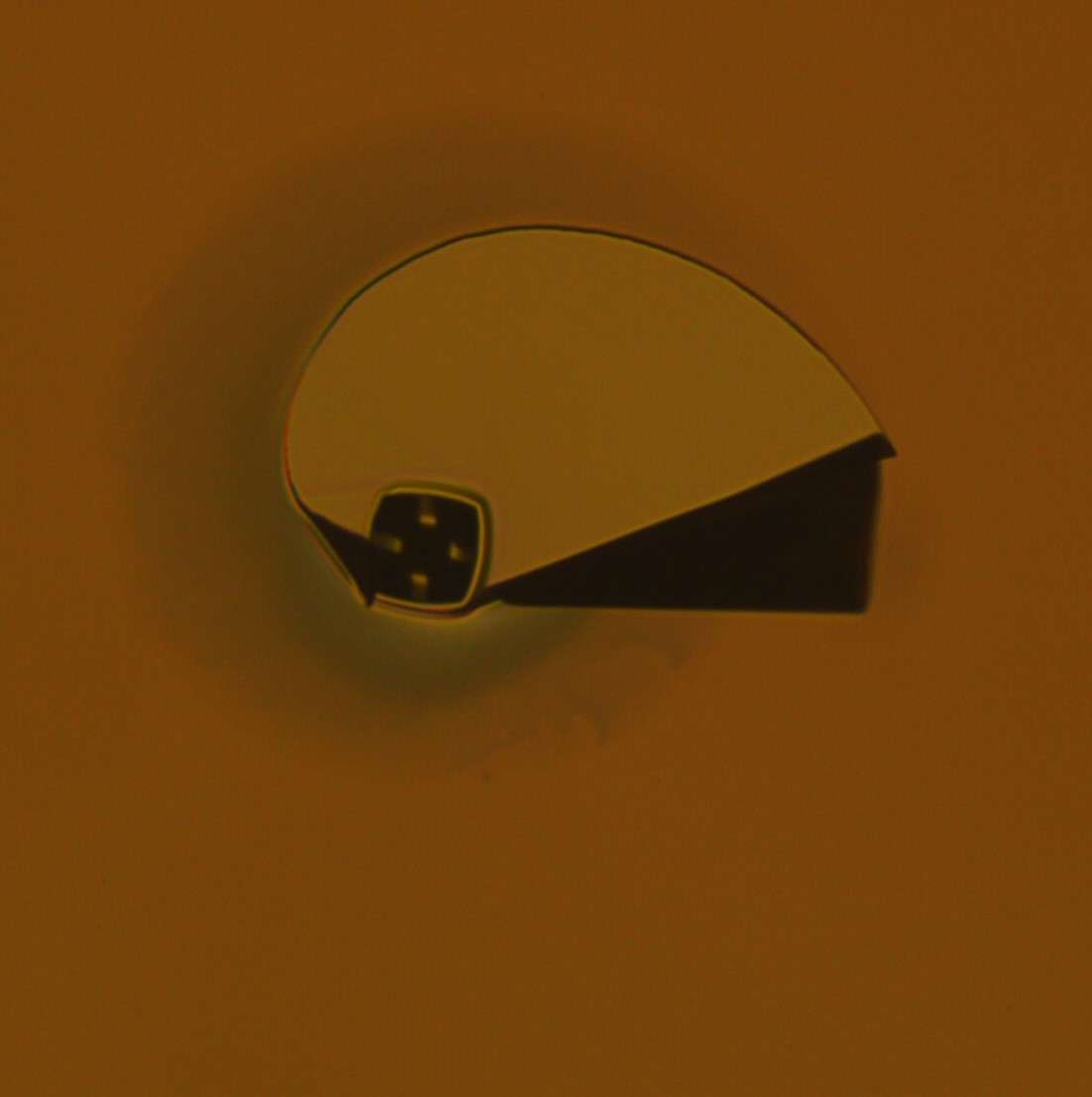}
	\includegraphics[width=2.0 in]{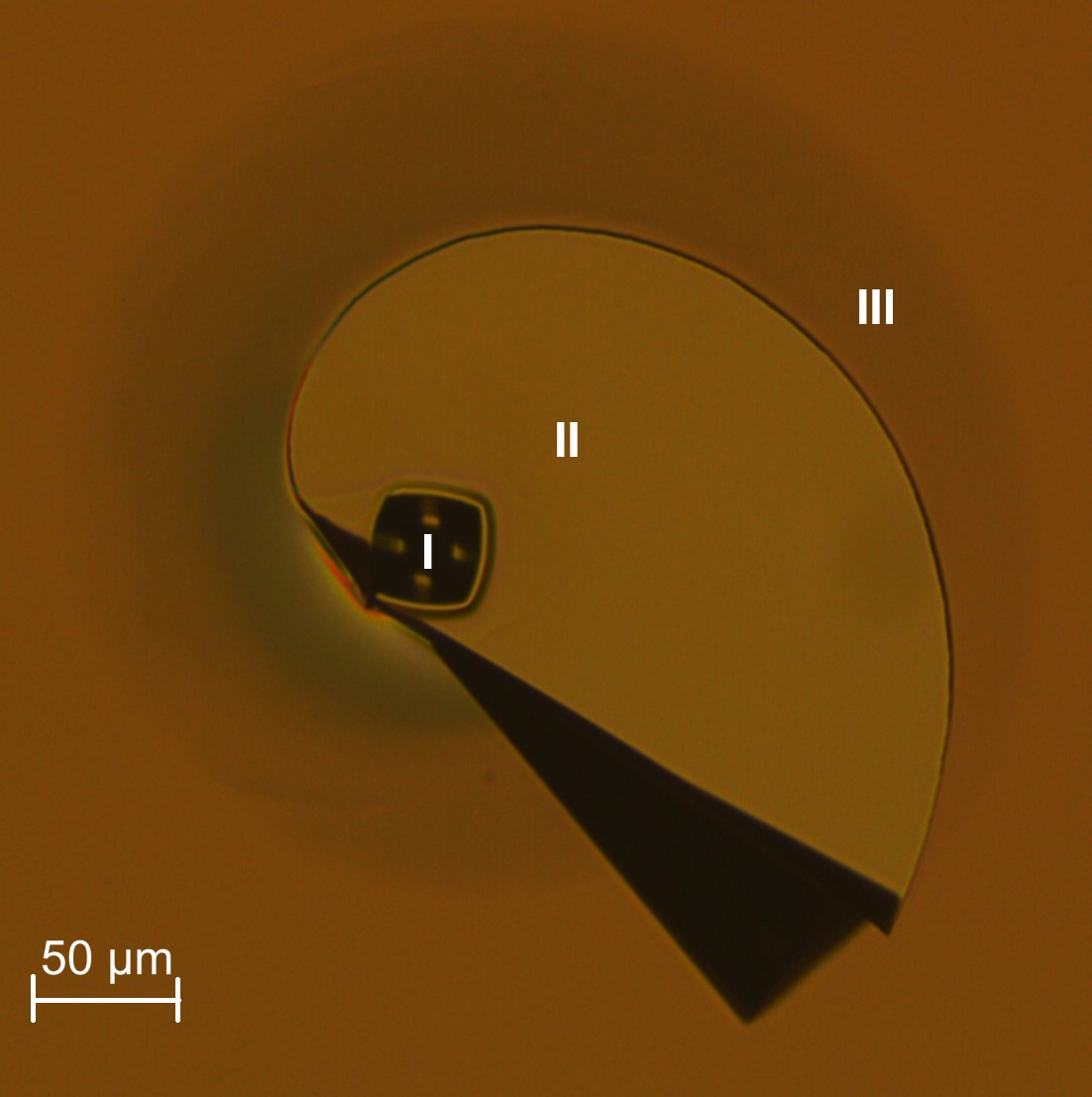}
	\caption{Three stills from the time lapse video of the growing Logarithmic spiral, taken at times 11 hrs, 21 hrs, 31 hrs 
	from the start of observations. Reflection microscopy with the sample under water. The growing crack in the metal film 
	defines the contour of the spiral. As it lifts from the surface, the metal film rolls up in a cone (dark area). The Ge-metal 
	contact line extends from the rim of the square etch pit at the center (region I) to the tip of the growing crack in the metal film, 
	sweeping out a spiral sector (region II). The darker halo outside the spiral (region III) is the delamination zone where 
	the metal film still covers the surface but etching solution has seeped in between the Ge surface and the film. }
	\label{fig:stills}
\end{figure}

\raggedbottom
\noindent 
Time lapse video microscopy of the growth of a one-arm Logarithmic spiral provides a window on how the etch pattern forms. 
The movie is available as Supp. Mat., while here we show a series of stills at successive times (Fig. \ref{fig:stills}). 
An initial instability of the metal film starting at the W side of the square etch pit develops into a growing crack in the film 
which eventually traces the outer edge of the spiral. There are actually two interacting cracks shaping the dynamics: 
the growing crack in the metal film and the propagating crack at the Ge-metal interface, which spans a straight line from the rim 
of the etch pit to the tip of the crack in the film. The metal film in the wake of the latter crack is detached from the Ge surface 
and ``rolls up'', exposing the bare Ge (visible as a lighter shade of ochra, while the rolled up metal film appears dark). 
As the outer crack in the metal film grows, tracing the contour of the Logarithmic spiral, the inner edge of the Ge-metal 
separation line remains pinned to the border of the etch pit and retraces this border. In this particular example, at later times, 
the pinning point detaches fom the rim of the etch pit and joins the line previously traced by the outer crack 
(which is the Ge-metal interface line; see movie in Supp. Mat.). The detached metal film rolls up due to residual stress in it: 
later we use the curvature of the rolled up metal shards to estimate the local residual stress in the film. \\ 
The Ge-metal contact line sweeps out a growing sector of the spiral as the pattern grows; the etching reaction is catalyzed at this 
front, i.e. along the Ge-metal contact line. Elsewhere there is either no etching solution in contact with the Ge (where the intact 
metal film protects the surface), or there is no catalyst (where the metal film is removed from the surface). The amount of material 
etched away from the Ge surface at one spot must be proportional to the dwell time of the metal-Ge contact line at that spot, or 
inversely proportional to the local normal velocity of the contact line (orthogonal to itself). That normal velocity is small close to 
the etch pit and large at the outer edge of the spiral: $v_n = r d \theta / dt$ where $r$ is the radial distance 
from the pit and $d \theta / dt$ the angular velocity at which the spiral rotates. Thus we expect the topographic profile $H(r)$ 
of the etched Ge surface to be sloping ``upwards'' in the radial direction as $H(r) \propto - 1 / v_n \propto - 1 / r$ , 
$r_{min} < r < r_{max}$ , where $r_{min}$ is the size of the etch pit and $r_{max}$ corresponds to the outer edge of the spiral. 
Later we discuss post mortem topographical maps of the Ge surface obtained with a profilometer which are consistent with this 
mechanism. 

\begin{figure}[H]
	\centering
	\includegraphics[width=2.4 in]{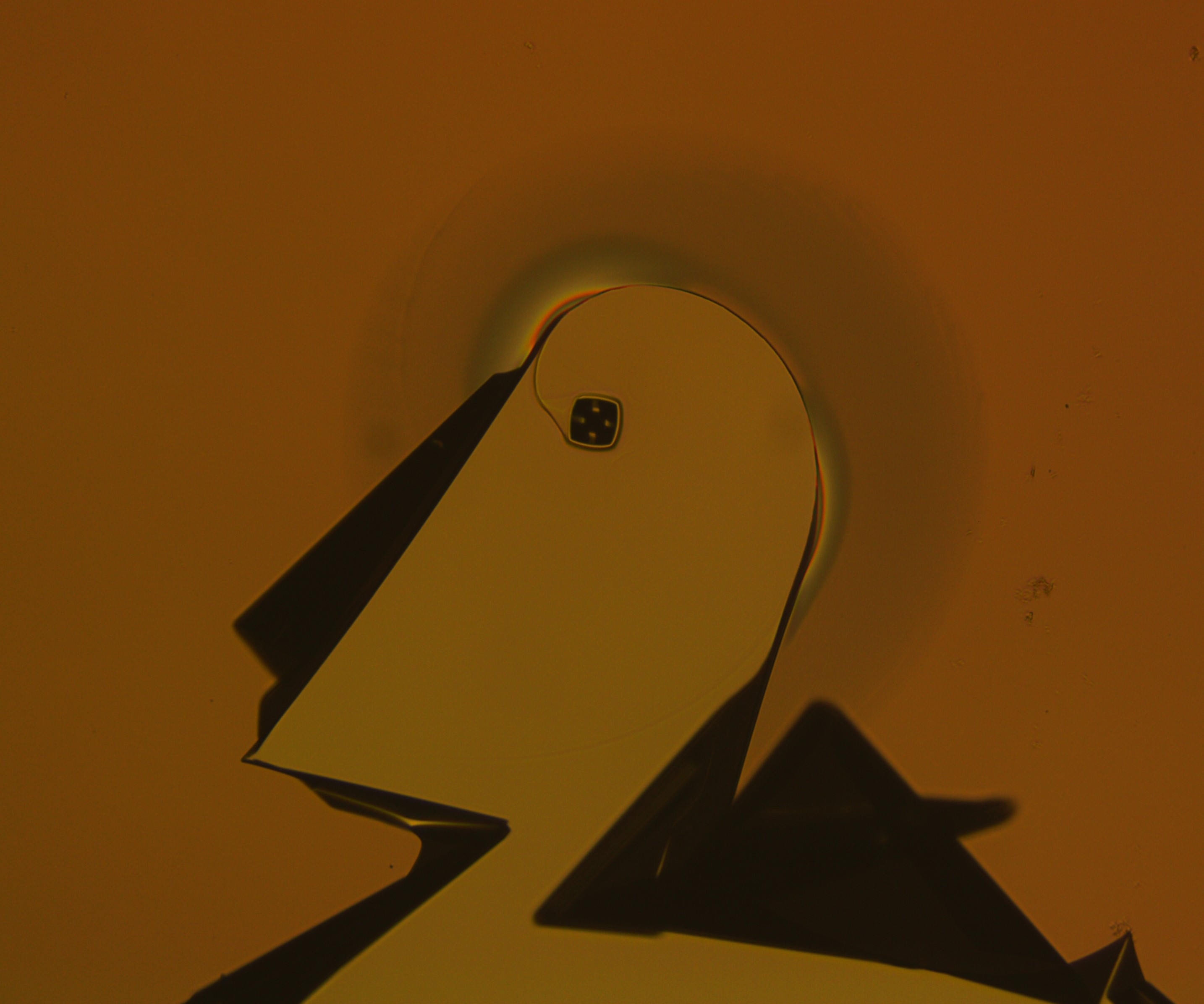}
	\includegraphics[width=2.4 in]{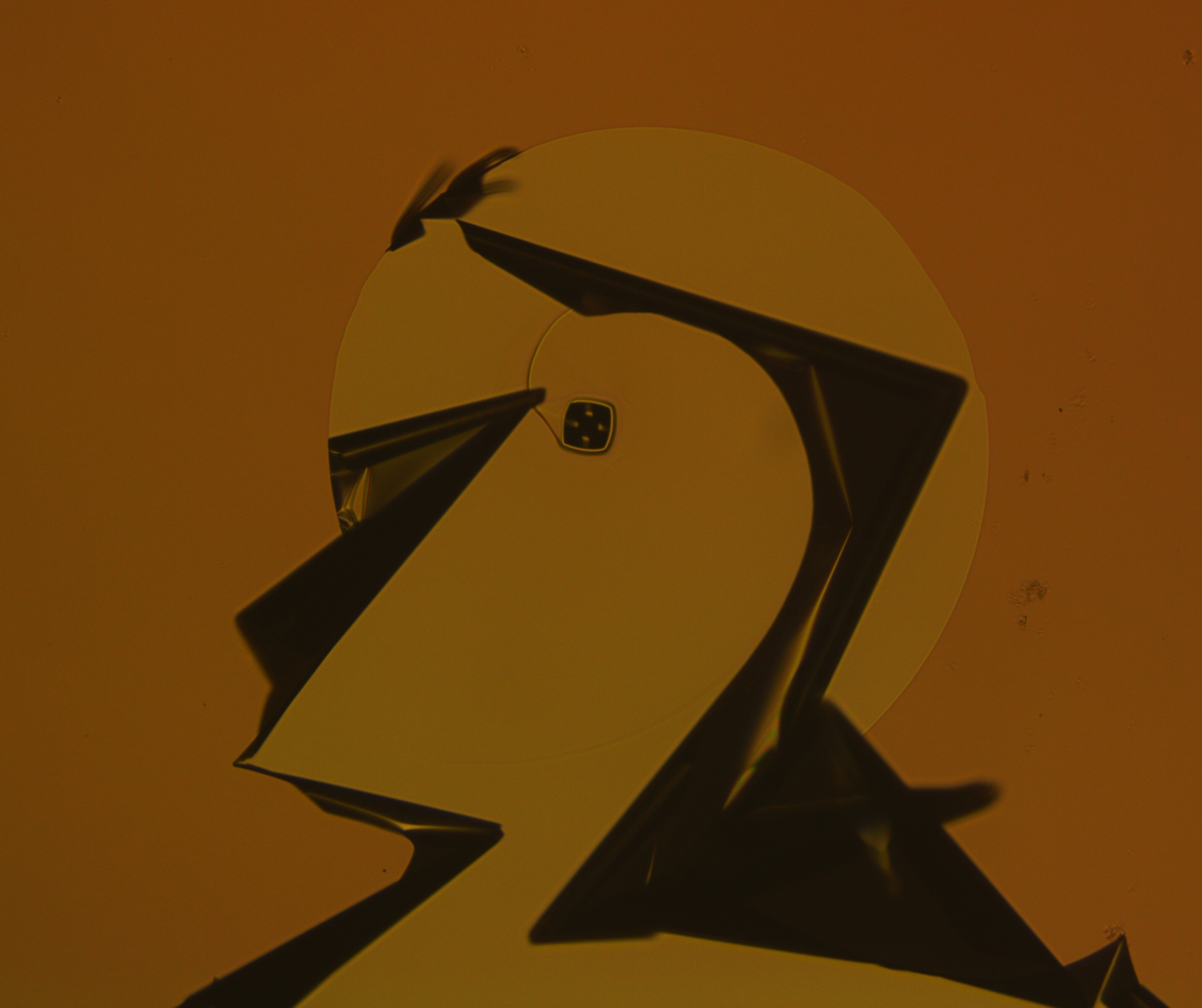}
	\caption{Two successive stills (3 min apart) from the time lapse video. They show the metal film ``suddenly'' lifting off 
	the delamination zone (darker halo in the upper picture, corresponding to region III of Fig. \ref{fig:stills}) and exposing 
	the bare Ge surface (lower picture). The metal film which was covering the delamination zone in the upper picture 
	is rolled up and appears dark in the lower picture. }
	\label{fig:rupture}
\end{figure}

Close inspection of the movie and the stills of Fig. \ref{fig:stills} reveals another component of the growth mechanism, visible as a radially 
expanding darker halo; the outer edge of the growing spiral meets this slightly darker expanding region at the tip of the growing crack 
in the metal film. We associate the halo region with delamination of the metal film from the Ge surface, a region where presumably 
a thin layer of etching solution has diffused in between the metal and the Ge. In this region, the etching reaction is active at the 
delamination front but presumably diffusion limited, as chemicals have to diffuse to and from the delamination front. On the other hand, 
the etching reaction at the exposed Ge-metal contact line is presumably rate limited. \\ 
The interpretation of the darker halo as a region of delamination of the metal film is confirmed later on in the movie 
($t \approx 1:16$  in the movie), when the metal covering this whole region ``suddenly'' lifts off (i.e. within one time step 
of the video lapse ($= 3 \, min$). We show this in Fig. \ref{fig:rupture} which displays two successive time steps of the movie, 
before and after this event. The rolled up metal film in the second picture was covering the delamination zone in the first, and 
the bare Ge exposed in the second picture corresponds to the delamination zone in the first.  

\begin{figure}
	\centering
	\includegraphics[width=2.0 in]{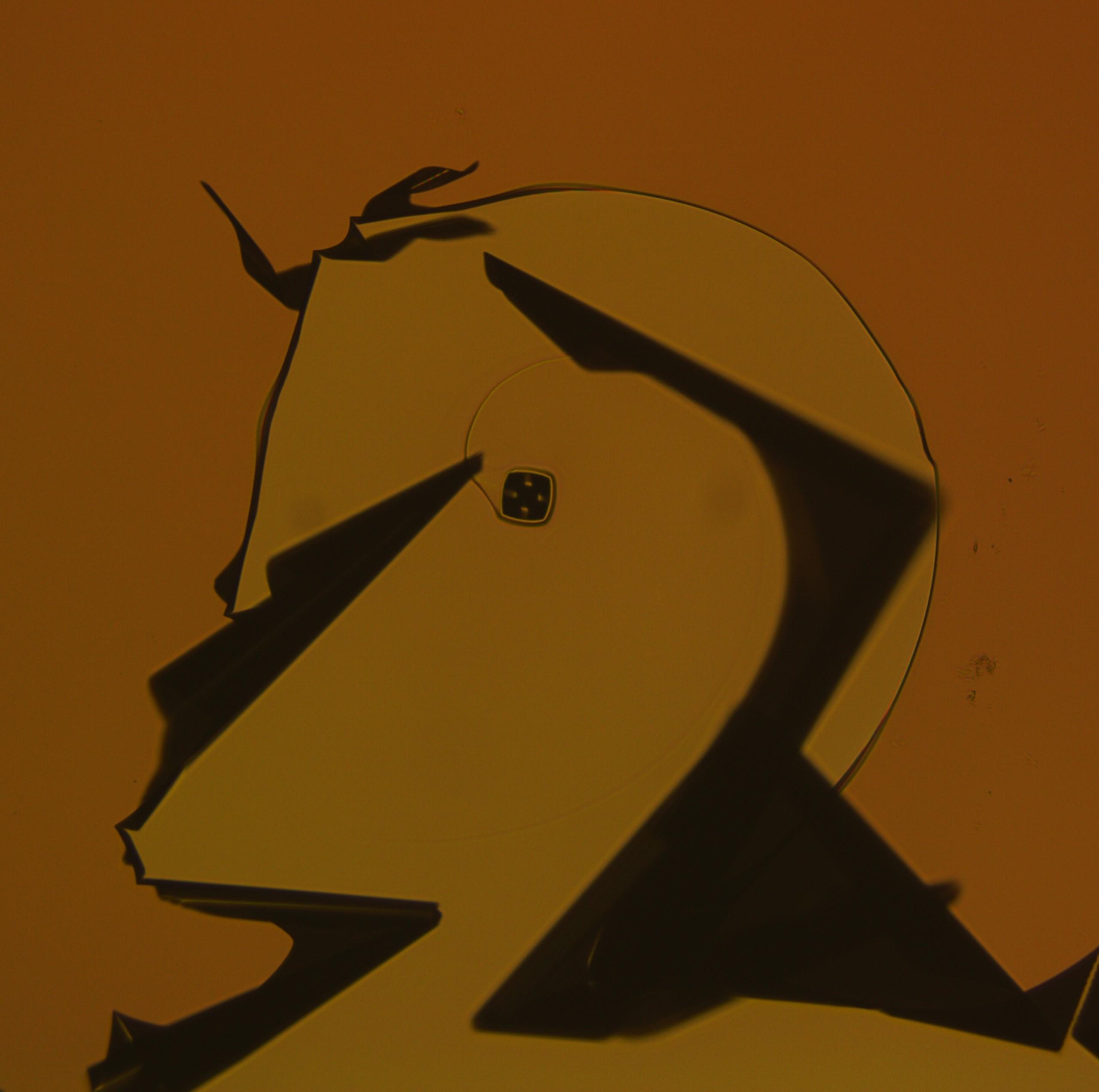}
	\includegraphics[width=2.0 in]{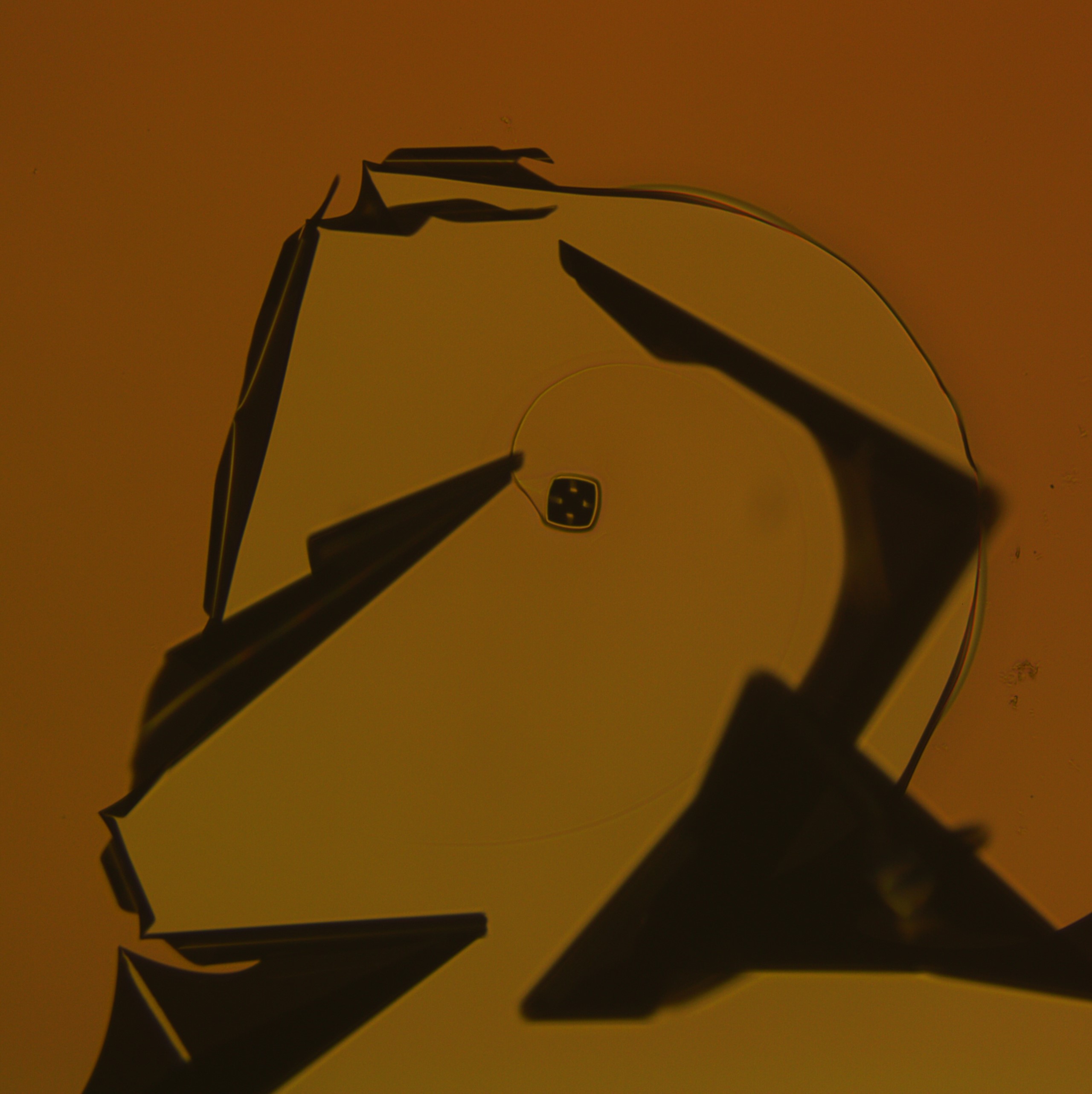}
	\caption{Two stills (5 hrs apart) from the time lapse video showing the transition to 
	the Lotus flower growth mode. The brighter region is the bare Ge, the deeper orange the metal film covered surface, and 
	the black parts are the rolled up metal film. In the NW corner we see a straight segment of Ge-metal contact line advancing, 
	while the metal film rolls up into a cylinder. }
	\label{fig:Lotus_growth_mode}
\end{figure}

Another noteworthy feature of the growth dynamics is visible in the movie at still later times, 
for growth occurring far from the etch pit. Fig. \ref{fig:Lotus_growth_mode} 
shows two stills for this regime: in the WNW direction we see a flat front 
propagating, that is, the metal-Ge contact line advances perpendicular to itself all at the same speed, i.e. with no rotation. 
This growth mode is involved in forming the ``lotus flower'' pattern. The same structure can transition from spiral close to 
the etch pit to ``lotus flower'' further away, as we see also in Fig. \ref{fig:Lotus_flower}. \\ 

\begin{figure}
	\centering
	\includegraphics[width=3.5 in]{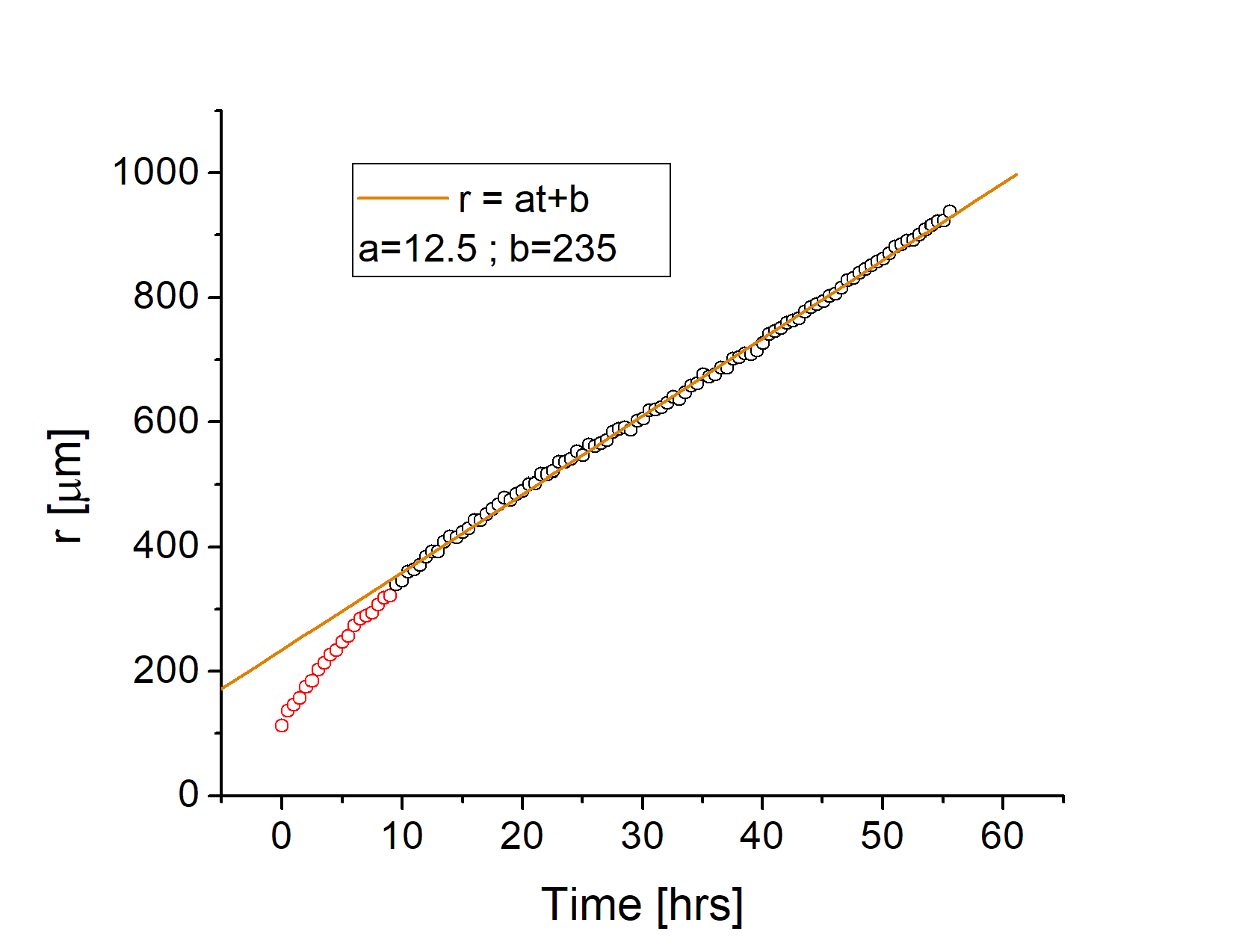}
	\caption{Radial position $r(t)$ of the tip of the crack in the metal film (see Fig. \ref{fig:stills}) in the course of time; 
	the origin is chosen at the center of the etch pit. After an initial transient ($0 < t < 9 \, [hrs]$; red symbols), 
	the radius of the spiral grows linearly in time. We attribute the initial transient to ``finite size effects'', being noticeable 
	when the spiral radius is comparable to the size of the etch pit. The solid line is a linear fit of the data for $t > 9 \, [hrs]$. 
	It extrapolates to $r = 0$ for $t = - 18.8 \, [hrs]$. The zero of time in the plot is the start of observations. }
	\label{fig:r_vs_t}
\end{figure}

\noindent {\bf Growth dynamics.} The time evolution  extracted from the time lapse video admits a relatively simple 
quantitative description, which we now present. 
Fig. \ref{fig:r_vs_t} is a plot of the measured radius $r$ of the spiral vs time. The radius is measured from the center of the etch pit 
to the outermost point of the growing spiral. That is, $r$ is the radial coordinate of the tip of the growing crack in the metal film, 
with the origin of the (polar) coordinate system at the center of the pit. The zero of time in the plot just corresponds to the beginning 
of observations. 

\begin{figure}[H]
	\centering
	\subfigure[\label{fig:theta_t}]{\includegraphics[width=3.5 in]{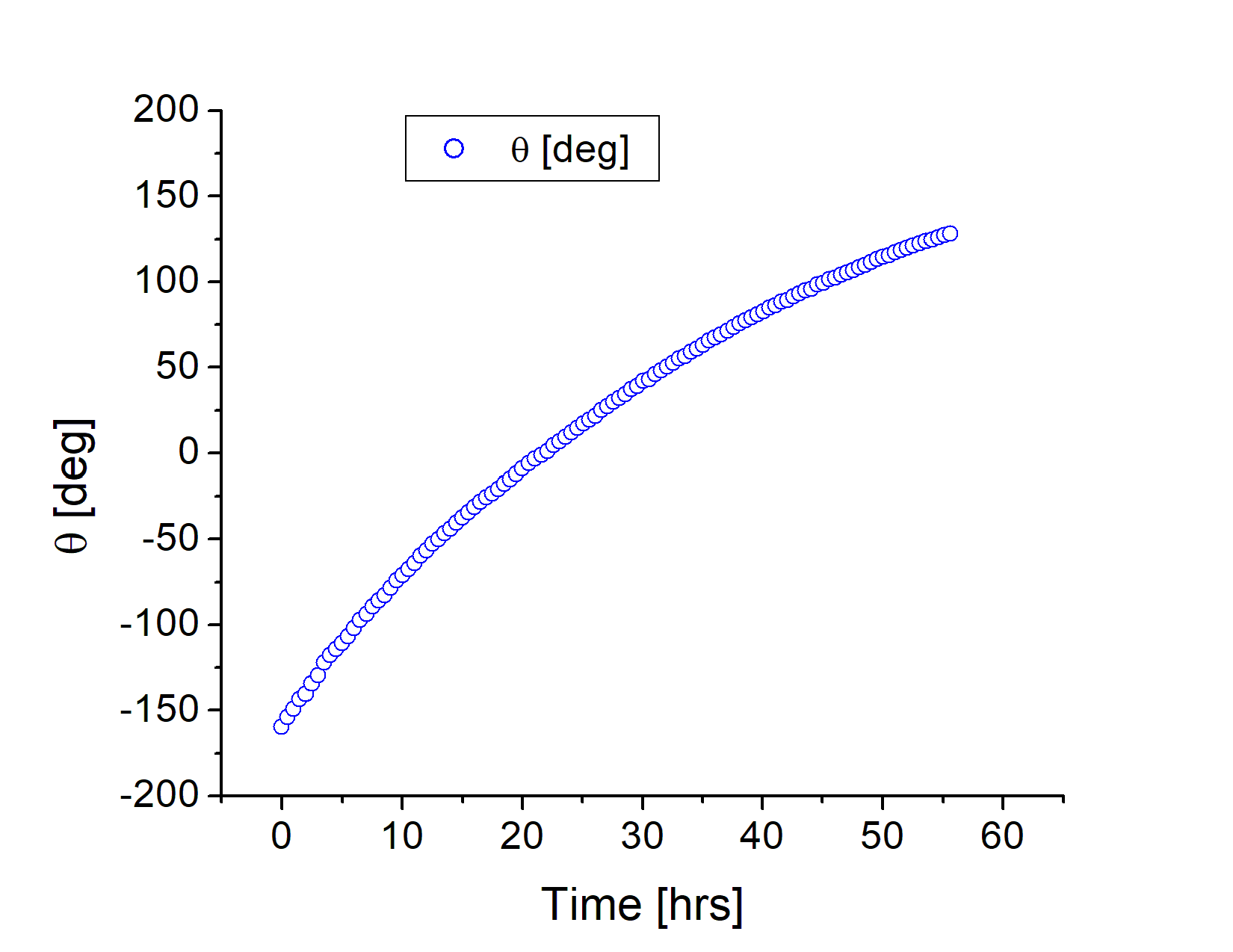}}
	\subfigure[\label{fig:theta_Log_t}]{\includegraphics[width=3.5 in]{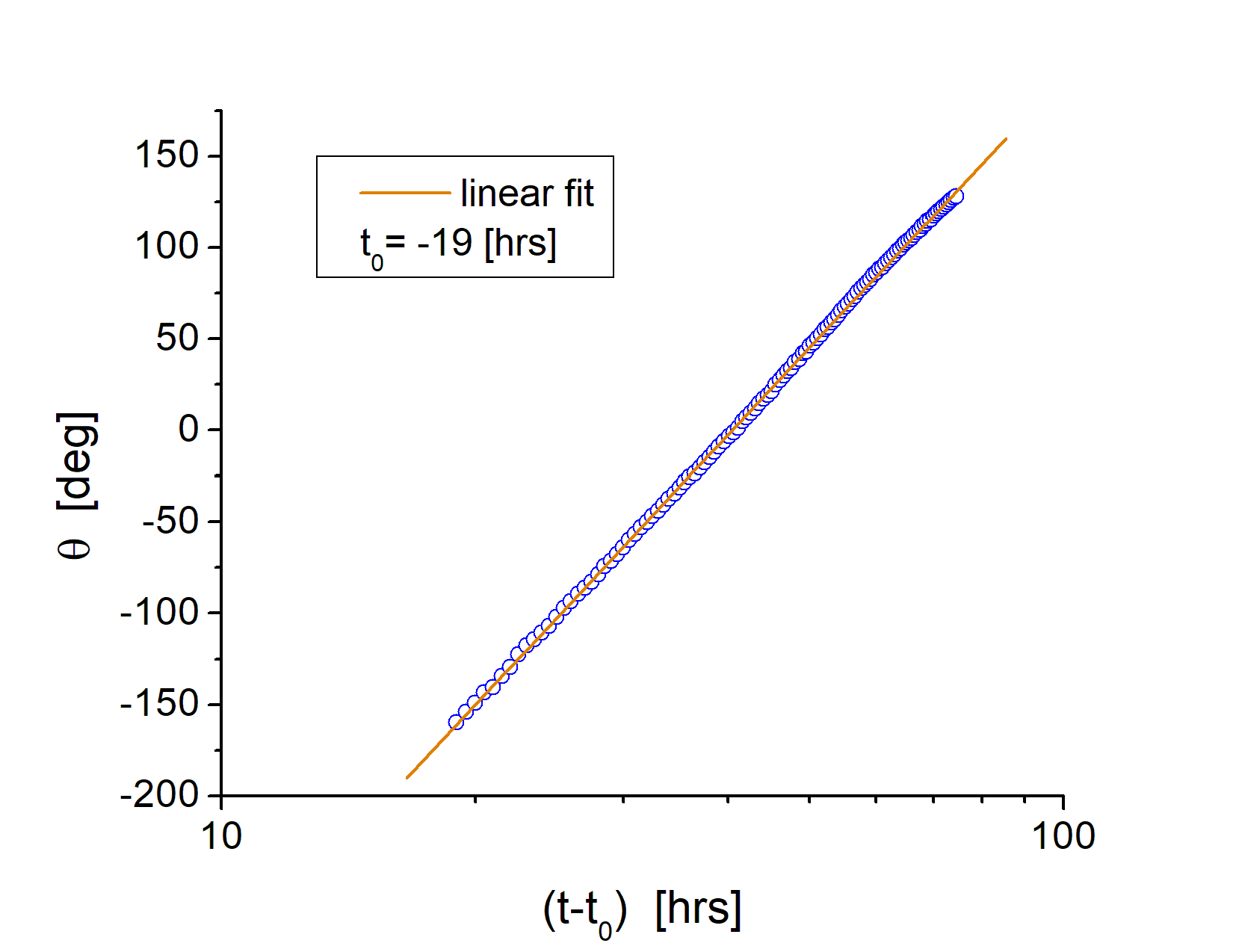}}
	\caption{(a) Polar angle $\theta (t)$ describing the position of the crack tip in the metal film which traces the outer contour 
	of the spiral. The angular velocity $d \theta / dt$ decreases as the spiral grows. \\ 
		(b) Same data as in (a) plotted vs $Log (t - t_0) $, where $t_0 = - 19 \, hrs$. The solid line is a linear fit. }
	\label{fig:theta}
\end{figure}

\noindent We see two phases in the growth: for $t > 9 \, hrs$ ($r > 300 \, \mu m$) the radius increases linearly in time ($dr / dt = const.$). 
At earlier times (corresponding to $100 < r < 300 \, \mu m$) $r(t)$ deviates from a straight line; later we argue that we may 
attribute this deviation to the finite size of the etch pit compared to $r$. The solid line in the figure is a linear fit of the measurements for 
$t > 9 \, hrs$ ; it gives a radial velocity $dr / dt \approx 12.5 \, \mu m / hr$ and it extrapolates to $r(t_0) = 0$ for $t_0 = - 18.8 \, hrs$. 
Fig. \ref{fig:theta_t} shows measurements of the angular coordinate $\theta (t)$ of the outer edge of the spiral, that is, $(r , \theta )$ are 
the polar coordinates of the tip of the growing crack in the metal film. 
The angular velocity $d \theta / dt$ decreases with time, approximately logarithmically. This is shown in Fig. \ref{fig:theta_Log_t} 
(same data as part (a)), where we plot $\theta$ vs $Log (t - t_0)$ , with $t_0 = - 19 \, hrs$. For this plot we adjusted the value 
of $t_0$ to make the data fall on a straight line (i.e. such that the plot is neither convex nor concave). Remarkably, this procedure 
returns the value $t_0 = - 20 \pm 2 \, hrs$ (for the plot of Fig. \ref{fig:theta_Log_t} we chose $t_0 = - 19 \, hrs$ to point out 
the consistency with the value extrapolated from the plot of $r(t)$ ). 
In summary, the trajectory of the crack tip, which traces 
the outer contour of the spiral, appears to emanate from a singularity at time $t_0 \approx -19 \, hrs$ from the start of observations. 
In polar coordinates $(r , \theta)$, and setting the origin of time at $t_0$ and the origin of space at the singularity (the center of the 
etch pit), the crack trajectory is 

\begin{equation}
	r(t) = a t \quad , \quad \theta (t) = \gamma \, log_{10} (t) + \theta_1 
	\label{eq:crack_trajectory}
\end{equation}

\noindent with $t$ in hrs, $\theta$ in degrees, $a = 12.5 \, \mu m / hr$ , $\gamma = 491$ , $\theta_1 = - 789$. 
We believe the singularity in question is the core of a screw dislocation perpendicular to the Ge surface; below we discuss 
measurements of the local strain in the metal film which are consistent with this statement. 

\begin{figure}[H]
	\centering
	\includegraphics[width=2.5 in]{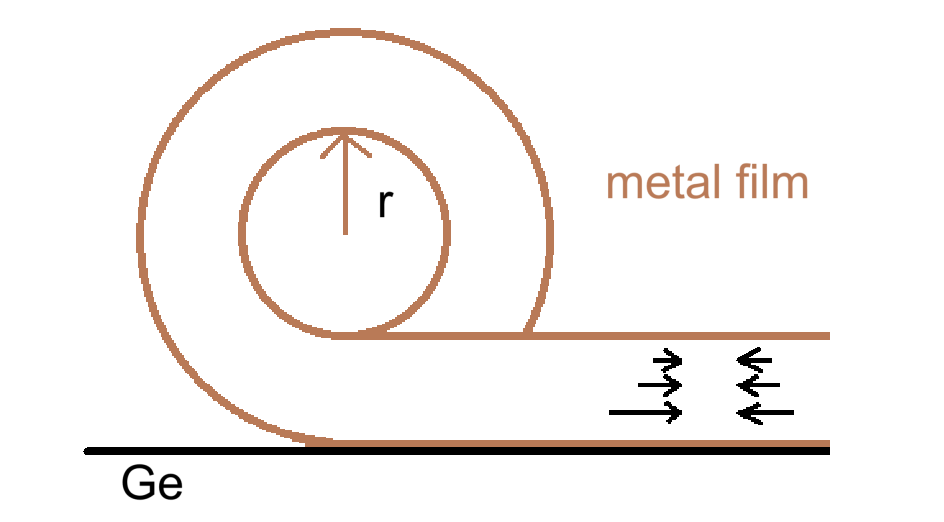}
	\caption{Sketch of the Ge-metal interface. The metal film is under compressive stress (arrows), and once detached 
	from the Ge surface it curls up.}
	\label{fig:Sketch}
\end{figure}

\noindent {\bf Strain measurements.} We see from the movie and the stills that as the spiral grows and the metal film detaches from 
the Ge surface, it rolls up. This indicates that while attached to the Ge surface, the metal film was under stress. After detaching 
from the surface, the film releases the stress by curling up. This phenomenon provides a way to measure the strain locally 
in the attached film. To make the argument, consider a uniaxial, compressive strain imparted by the Ge surface on the metal film interface 
(Fig.  \ref{fig:Sketch}). Assume the detached film, in mechanical equilibrium, rolls up in a cylinder of (inner) radius $r$. 
The outer radius is then $(r + h)$ where $h$ is the thickness of the film ($h << r$). If we imagine to ``unroll'' the cylinder back 
on a flat plane, the free surface of the film (which is stress-free) will have length $\ell = 2 \pi r$, while the film surface in contact 
with the Ge must shrink from $L = 2 \pi (r + h)$ to $\ell$, corresponding to a strain $\epsilon = (L - \ell) / \ell = h / r$ 
where $h$ is the film thickness and $r$ the radius of curvature of the rolled-up film. 

Typical radii of curvature we observe 
for the rolled up film are of order $r \approx 10 \, \mu m$ ; for $h = 20 \, nm$ this gives typical residual strains of order 
$\epsilon \approx 2 \times 10^{-3}$ , or residual stresses of order $\sigma = Y \epsilon \approx 540 \, MPa$ where 
$Y \approx 270 \, GPa$ is the Young modulus for (bulk) Cr. However, the rolled up film in the stills of Fig. \ref{fig:stills} 
forms cones rather than cylinders. The same analysis (by unfolding the cone into a sector of a circle etc.) reveals that 
the residual strain in the film has an azimuthal component which is: 

\begin{equation}
	\epsilon_{\theta} = \frac{1}{sin (\alpha / 2)} \, \frac{h}{r}
	\label{eq:film_strain}
\end{equation}

\noindent where $\alpha$ is the opening angle of the cone and $r$ the radial distance from the (extrapolated) vertex of the cone. \\ 
Remarkably, the strain field just described is also the strain field due to a screw dislocation (in an infinite isotropic medium). 
For the latter (see e.g. \cite{GI_Taylor1934}), with the z axis along the dislocation core, the only non zero components 
of the strain tensor are: 

\begin{equation}
	u_{xz} = u_{zx} = \frac{b}{4 \pi} \, \frac{- sin \theta}{r} \quad , \quad u_{yz} = u_{zy} = \frac{b}{4 \pi} \, \frac{cos \theta}{r}
	\label{eq:dislocation_strain}
\end{equation}

\noindent where $\theta$ is the angle in polar coordinates in the xy plane and $r$ the distance from the z axis; $b$ is the Burgers vector 
of the dislocation. In the experiments, we measure essentially this same strain field at the Ge - metal interface in a region 
around a center (marked by the etch pit) which will give rise to the spiral pattern. We say ``essentially'' because in the experiments, 
due to the finite size of the etch pit region, the coordinates $(r , \theta)$ are not as precisely defined as in theory; specifically, 
the extrapolated vertex of the cone formed by the rolled up film is not exactly the center of the etch pit, but rather turns around the 
outer primeter of the etch pit as the spiral turns (the metal film being apparently pinned to the perimeter of the etch pit). 

\begin{figure}[H]
	\centering
	\includegraphics[width=3.5 in]{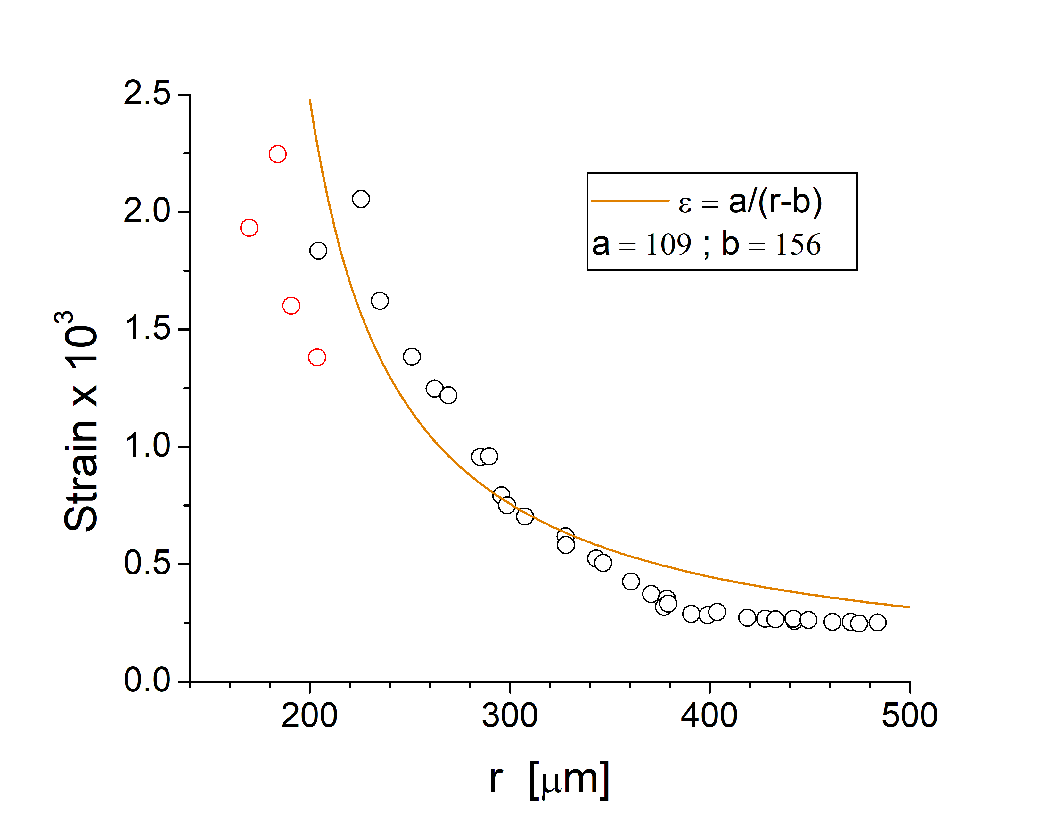}
	\caption{Residual azimuthal strain $\epsilon_{\theta}$ in the metal film vs $r$ measured for the spiral pattern of Fig. \ref{fig:stills} 
	using the rolled up metal film shapes (see text). $r$ is the distance between the apex and the base of the cones, and the strain 
	is calculated using eq. (\ref{eq:film_strain}). The solid line is a fit with the form $\epsilon_{\theta} = a / (r - b)$ 
	with $a = 109$ and $b = 156$. The 4 red data points at small $r$ have been excluded from the fit. }
	\label{fig:strain}
\end{figure}

In Fig. \ref{fig:strain} we plot measurements of the azimuthal strain $\epsilon_{\theta}$ obtained from the movie stills as 
in Fig. \ref{fig:stills}. $r$ is the distance between the apex and the base of the cones, measured on the images, 
and the strain is calculated using the measured opening angle of the cones (see Mat. \& Met.) according to (\ref{eq:film_strain}). 
These measurements show (or more conservatively, are consistent with the notion) that the residual strain in the metal film 
does decrease with distance from the etch pit, with roughly a $1 / r$ dependence. However, for the reasons stated above 
(basically the finite size of the etch pit), to fit the data we have to introduce a shift of origin for the distance $r$, which takes into 
account in an average way the motion of the apex of the cones around the rim of the etch pit (Fig. \ref{fig:stills}). The solid line 
in Fig. \ref{fig:strain} shows the function $\epsilon = a / (r - b)$, with $a = 109 \, [nm]$, $b = 156 \, [\mu m]$ obtained by fitting 
the data points, excluding the 4 red points at small $r$ ($r < 200 \, \mu m$), which seem to be very obviously affected by 
finite size effects.  
However, if we disregard these complications, the overall picture we propose is that an underlying screw dislocation in the Ge 
imparts a stress field at the Ge - metal interface which is one important component for the growth dynamics of the crack 
in the metal film and the Logarithmic spiral pattern. \\

\begin{figure}[H]
	\centering
	\subfigure[\label{fig:three_arm_pic}]{\includegraphics[width=2.5 in]{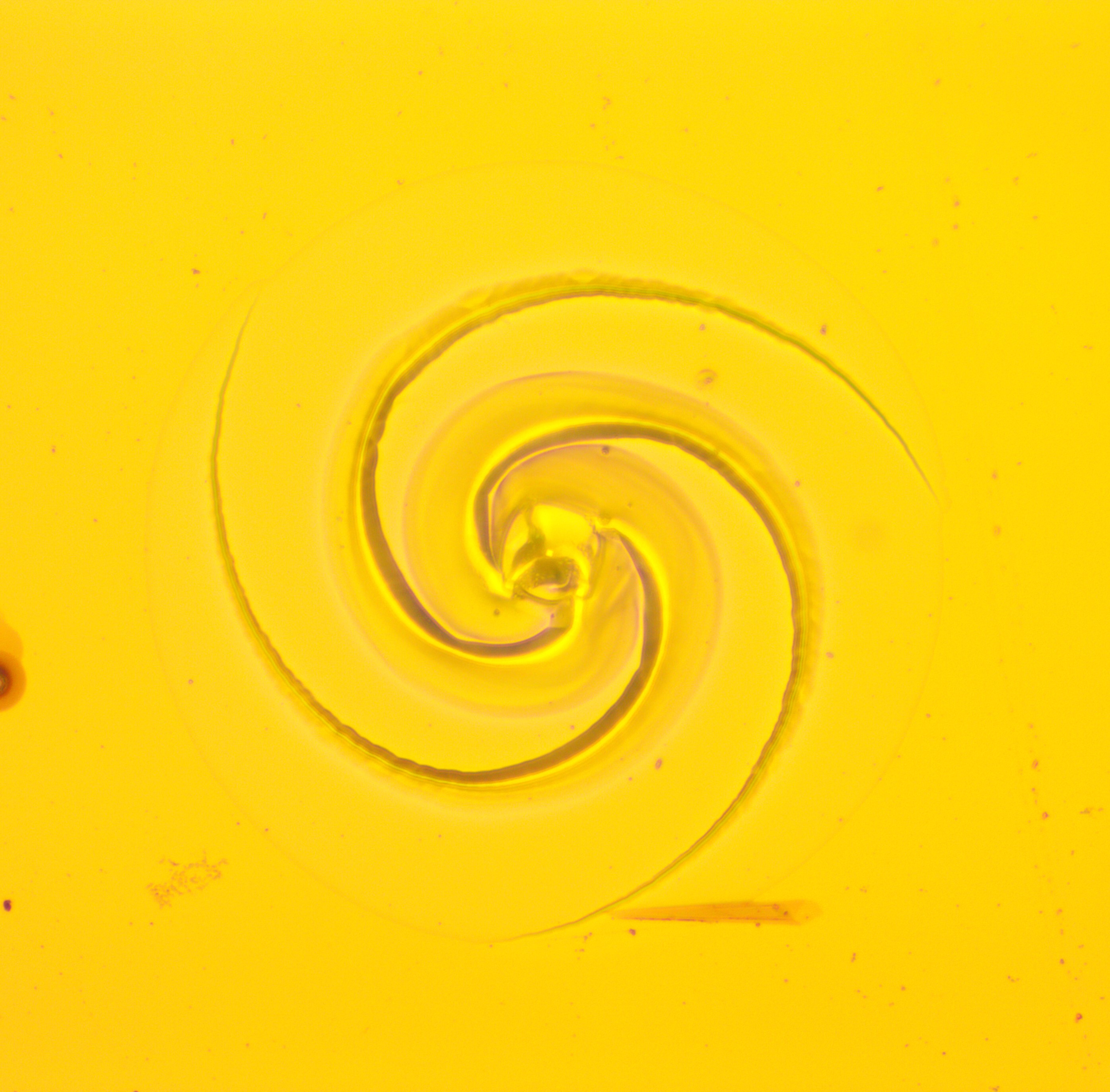}}
	\subfigure[\label{fig:three_arm_profile}]{\includegraphics[width=3.5 in]{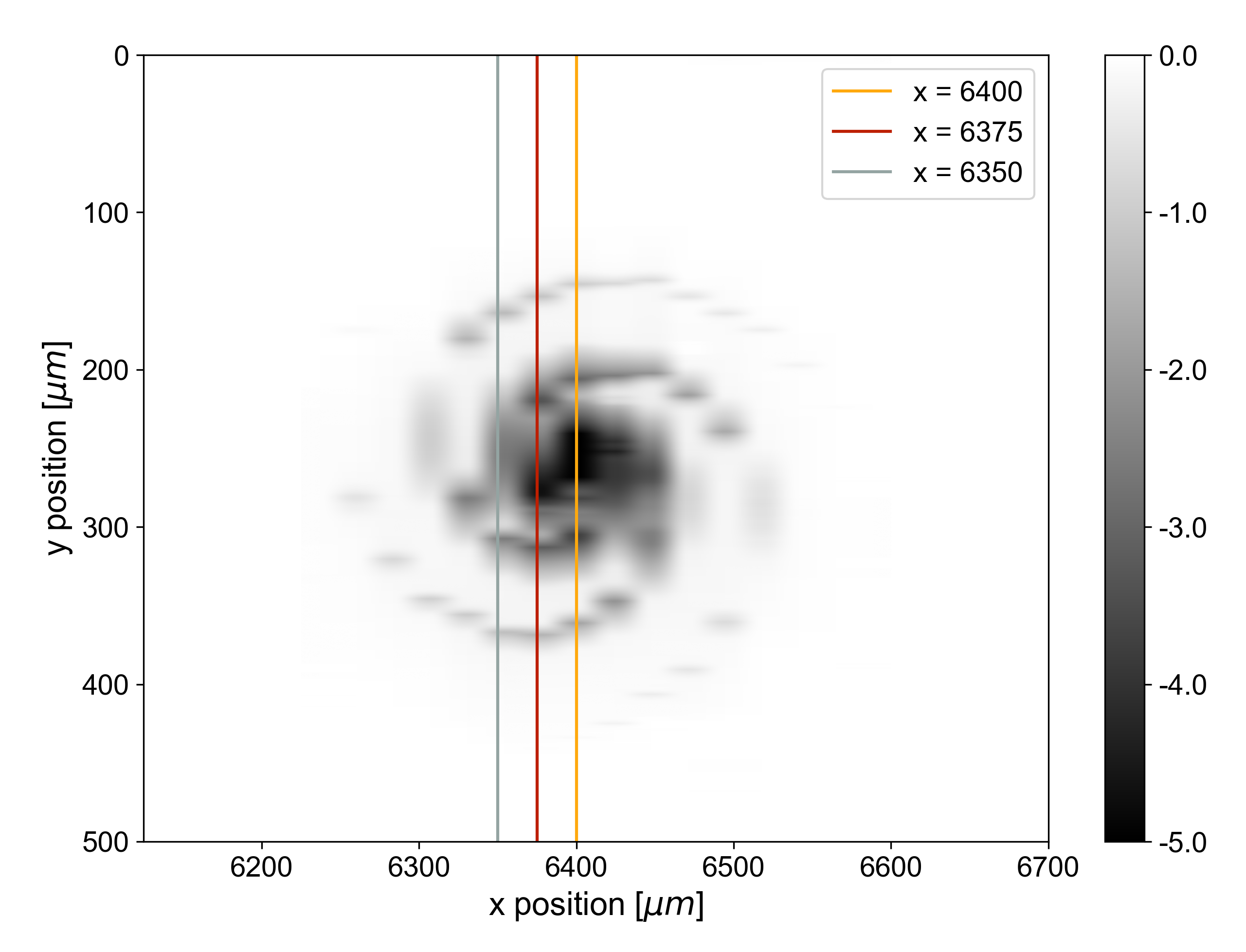}}
	\caption{(a) Light microscopy picture of a three-arms spiral (dry sample). As always, the pattern emanates from the square 
	etch pit at the center.  \\ 
		(b) Composite profilometer map of the surface topography of the pattern in (a). This 2D topographic map is constructed 
		from $x = const.$ line scans taken every $25 \, \mu m$. The grey scale shows the depth in $\mu m$. 
		The spiral arms are revealed as deeper grooves into the Ge surface. The three vertical lines mark the position of the line scans 
		shown in Figs. \ref{fig:Profile_6400B}, \ref{fig:Profile_75_50}. }
	\label{fig:three_arms}
\end{figure}

\noindent {\bf Topography of the etched Ge surface.} The optical microscopy pictures such as Fig. \ref{fig:Log_spiral} reveal 
the beautiful overall pattern but may be tricky to interpret in terms of the engraving depth of the etched Ge surface: 
grooves may look like ridges and vice versa. We therefore measured the actual surface profile for a few patterns by means of a 
profilometer. Fig. \ref{fig:three_arm_pic} shows the optical microscopy picture of a three-arms spiral, and Fig. \ref{fig:three_arm_profile} 
the corresponding topographic map where the grey scale represents surface depth $H$ below the unetched level $H = 0$. 
This map is a composite of vertical ($x = const.$) scans taken every $25 \, \mu m$. We recognize the $\sim 20 \, \mu m$ size 
central well (etch pit), which is about $6 \, \mu m$ deep (see below). The spiral arms, which might look like ridges in the optical 
micrograph, are in fact deep grooves. Following any one arm moving outwards from the central pit (i.e. turning clockwise) 
the Ge floor is sloping upwards towards the level of the unetched surface, and the same is true for the regions in between the arms. 
More detail is seen in the individual line scans. 

\begin{figure}
	\centering
	\subfigure[\label{fig:Profile_6400}]{\includegraphics[width=3.5 in]{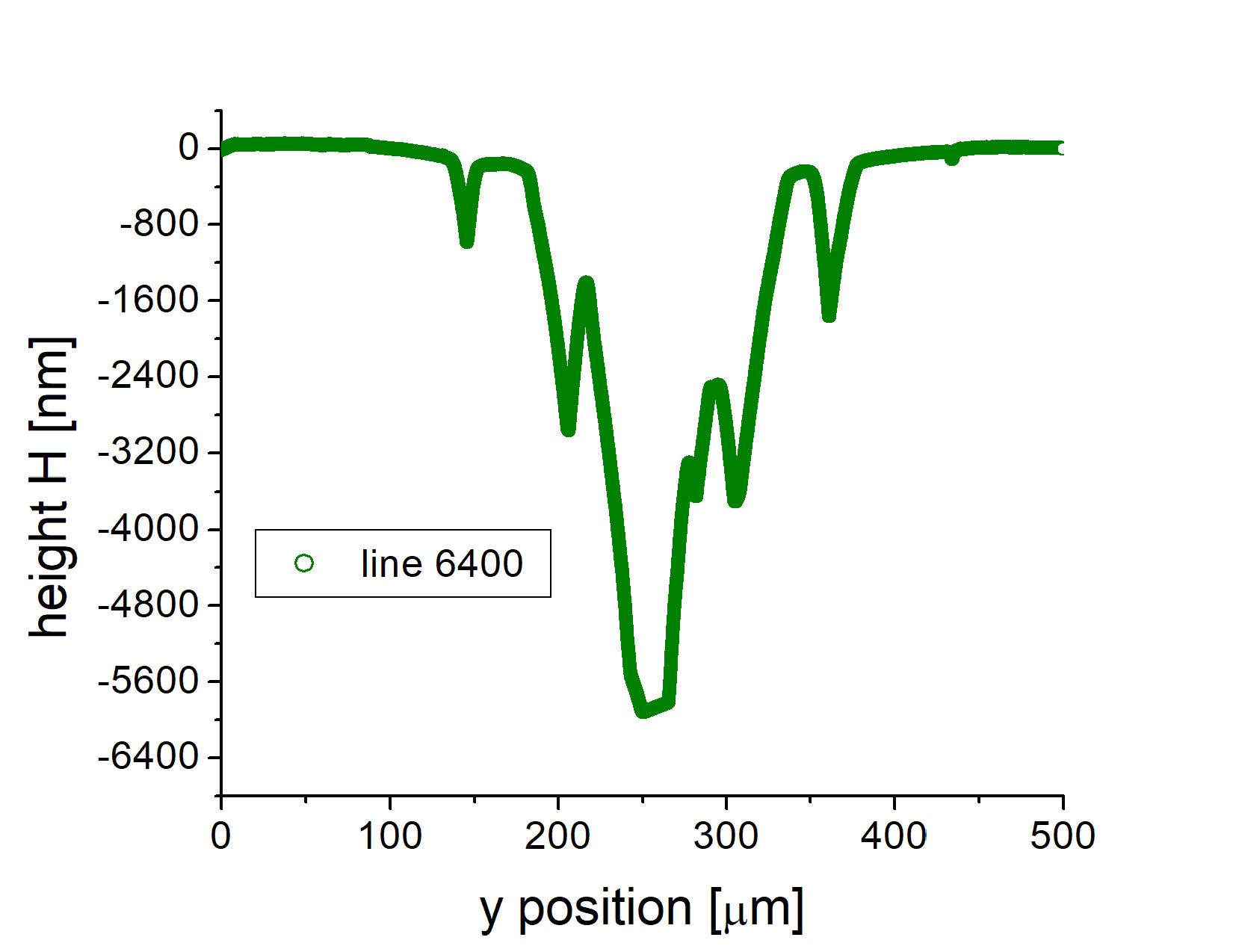}}
	\subfigure[\label{fig:P_6400_detail}]{\includegraphics[width=3.5 in]{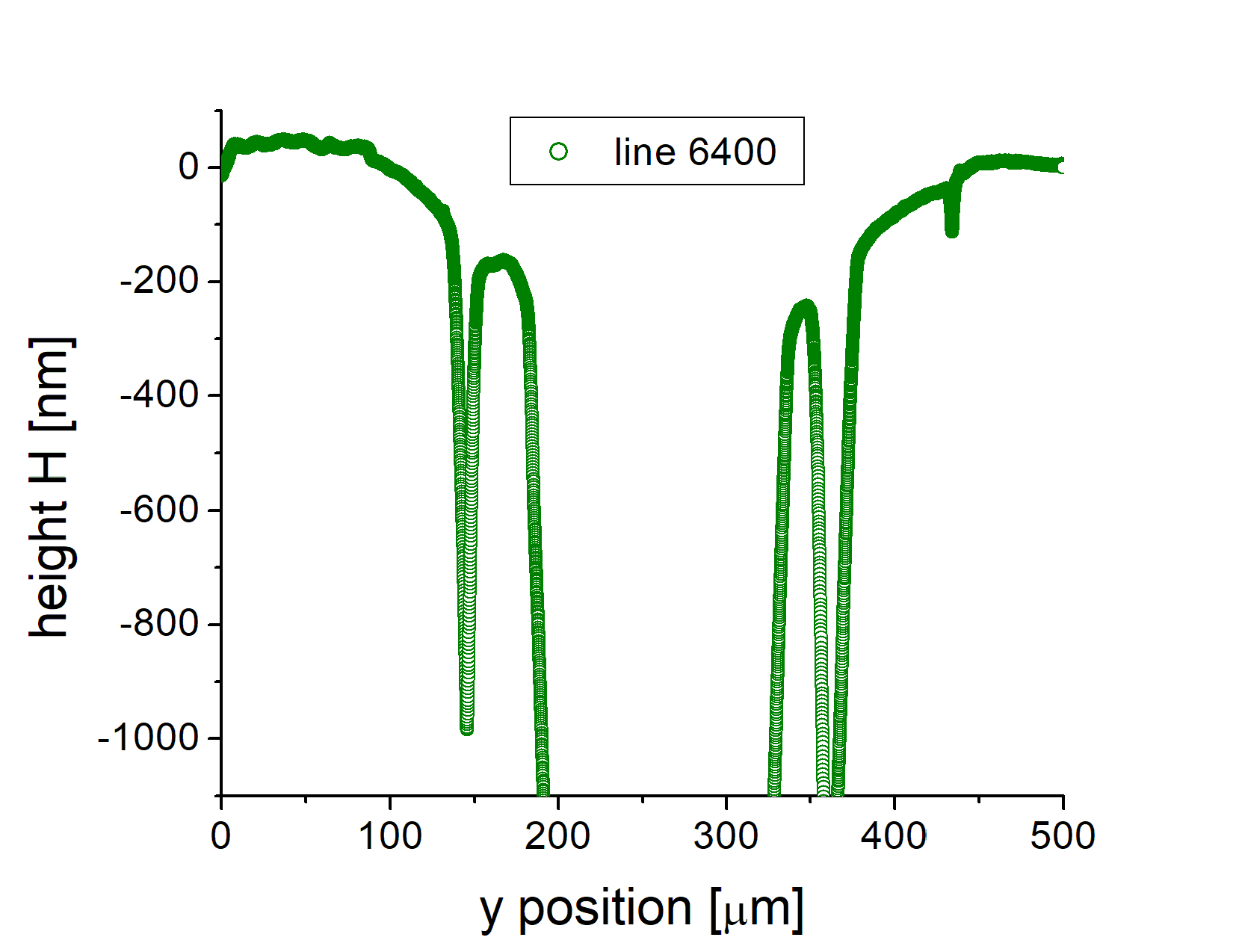}}
	\caption{(a) Profilometer scan of the line $x = 6400$ in Fig. \ref{fig:three_arm_profile}. The height is in $nm$ and $H = 0$ is the 
	level of the unperturbed surface. The central etch pit has a depth of $\approx 6 \, \mu m$ and the spiral arms are clearly seen 
	as the V-shaped grooves on either side of the etch pit (compare the y-position with Fig. \ref{fig:three_arm_profile}).   \\ 
	(b) Detail of the plot in (a). Careful examination shows that the plateau regions in between arms are not flat, but rather present a slope. }
	\label{fig:Profile_6400B}
\end{figure}

Fig. \ref{fig:Profile_6400} shows the line scan corresponding to the (vertical) line $x = 6400$ in Fig. \ref{fig:three_arm_profile}. 
Plotted is the height $H$ of the sample surface vs the $y$ coordinate (which is pointing downwards in Fig. \ref{fig:three_arm_profile}). 
The prominent features are the spiral arms and of course the central etch pit. The arms of the spiral are seen to correspond to deep 
asymmetric grooves (compare with Fig. \ref{fig:three_arm_pic}). For the line $x = 6400$, the upper spiral arm is crossed at 
$y = 146 \, [\mu m]$ and has a depth $H = - 905 \, [nm]$ below the unetched surface; the next arm is crossed at 
$y = 205 \, [\mu m]$ and its maximum depth is $H = - 2870 \, [nm]$. Then comes the etch pit, with a nearly flat floor 
($250 < y < 265 \, [\mu m]$) at $H = - 5850 \, [nm]$. For larger $y$ we see the other arms corresponding to the lower portion 
of the pattern in Fig. \ref{fig:three_arm_pic}. 

\begin{figure}
	\centering
	\subfigure[\label{fig:Profile_6375}]{\includegraphics[width=3.5 in]{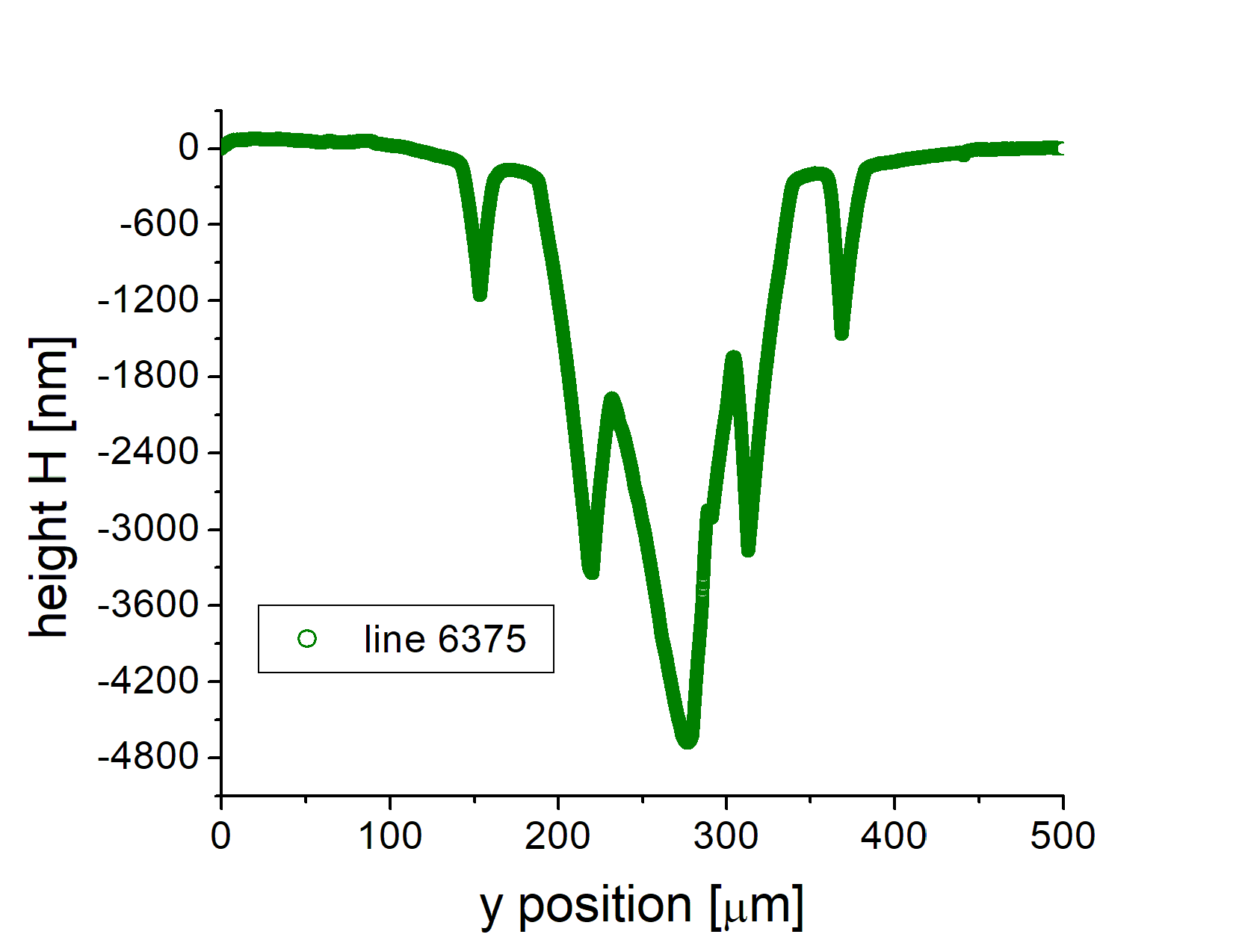}}
	\subfigure[\label{fig:Profile_6350}]{\includegraphics[width=3.5 in]{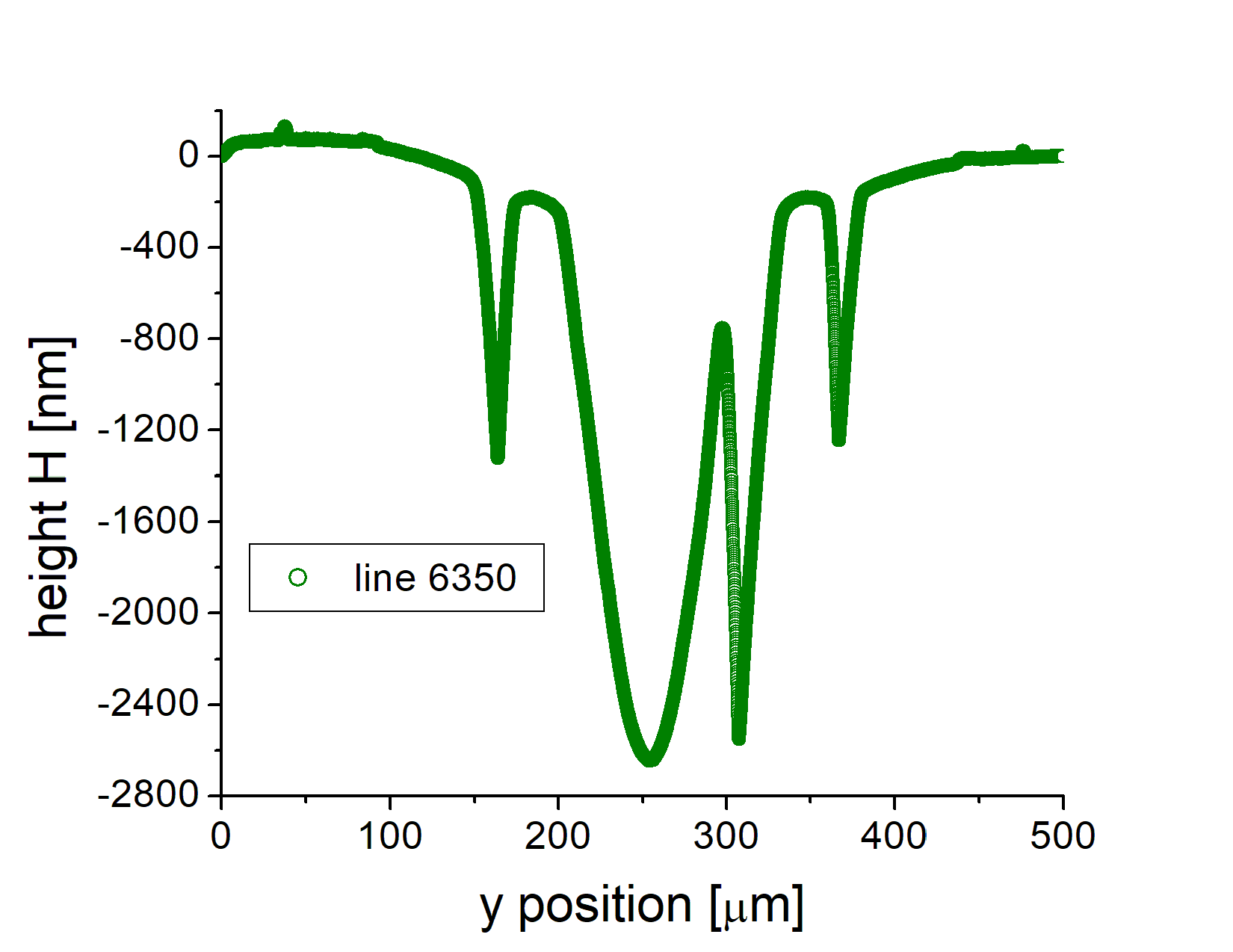}}
	\caption{(a) Profilometer scan of the line $x = 6375$ (red line in Fig. \ref{fig:three_arm_profile}).  \\ 
			(b) Profilometer scan of the line $x = 6350$ (grey line in Fig. \ref{fig:three_arm_profile}). }
	\label{fig:Profile_75_50}
\end{figure}

Closer examination provides further detail: for ($0 < y < 85 \, [\mu m]$) we see the flat, unetched surface. The region 
($85 < y < 130$) corresponds to the ``delamination zone'', visible as the circular halo above the upper spiral arm in Fig. \ref{fig:three_arm_pic}. 
This region slopes downwards by approximately $\Delta H = 90 \, [nm]$ for $\Delta y = 45 \, [\mu m]$ (Fig. \ref{fig:P_6400_detail}). 
After the deep groove (centered at $x = 145$) of the spiral arm, there is a $\sim 14 \, \mu m$ wide nearly flat plateau 
($156 < y < 170$ , $H = - 165$) corresponding to the lighter region in between the two upper spiral arms in Fig. \ref{fig:three_arm_pic}. 

\begin{figure}[H]
	\centering
	\subfigure[\label{fig:one_arm_pic}]{\includegraphics[width=2.5 in]{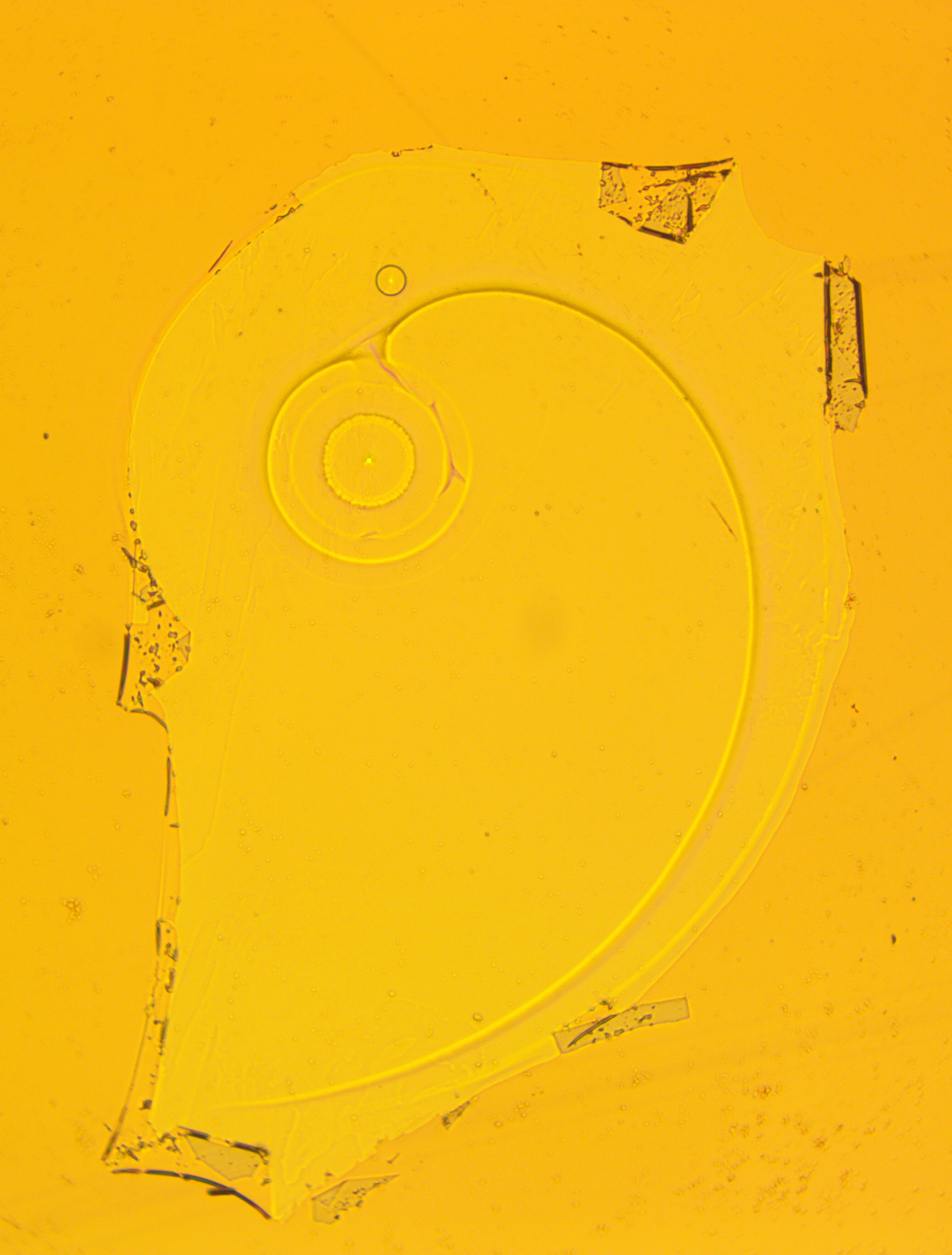}}
	\subfigure[\label{fig:one_arm_profile}]{\includegraphics[width=3.5 in]{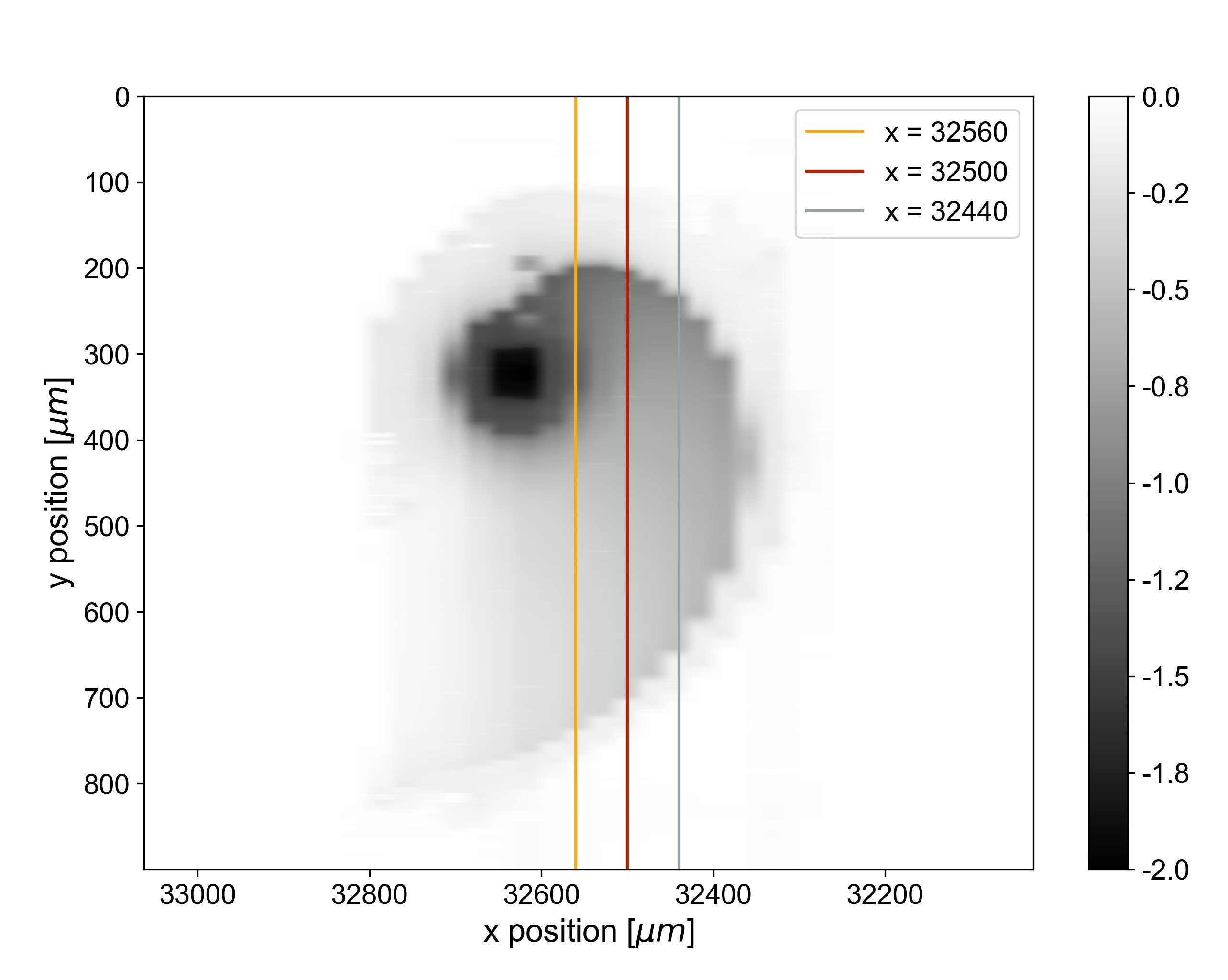}}
	\caption{(a) Light microscopy picture of a one-arm spiral. Dry sample exposing the bare Ge surface (lighter area) where the 
	metal film has lifted off. We use this simpler pattern to build a quantitative understanding of the etching process. \\ 
		(b) Composite profilometer map of the surface topography of the spiral in (a). The 2D map is constructed from $x = const.$ 
		line scans taken every $30 \, \mu m$. The grey scale represents the depth in $\mu m$. }
	\label{fig:one_arm}
\end{figure}
 
 Next comes the steep groove of the spiral arm (centered at $x = 205$), the surface sloping 
downwards by $\Delta H = 2700 \, [nm]$ for $\Delta y = 25 \, [\mu m]$ (a $\sim 10 \%$ incline: the reader is cautioned that when we say 
``steep'', we mean {\it steeper} than other parts of the engraved surface). This pattern is repeated mirror-fashion on the other side 
of the structure ($360 < y < 500$). 
In Fig. \ref{fig:Profile_75_50} we show two further profilometer cuts of the same structure, moving away from the central etch pit 
($x = 6375$ and $x = 6350$). The reader can easily relate the profiles to the pattern of Fig. \ref{fig:three_arms}. 
 The profilometer measurements can be interpreted quantitatively following one reasonable assumption; the corresponding model 
clarifies the mechanism leading to the observed engraving. For simplicity, we detail this analysis for a one-arm spiral. 
Fig. \ref{fig:one_arm_pic} shows the light microscopy picture of the dry sample; as before, the lighter colored area is bare Ge 
while the surrounding darker area is the metal film. The spiral shaped line corresponds to basically a step (``downwards'' 
traversing the line from the periphery towards the interior of the structure). The genesis of this line is seen (for a different spiral) 
from the stills of Fig. \ref{fig:stills}: it corresponds to the path of the crack in the metal film which defines the spiral sector swept by 
the metal-Ge contact line as the patterns grows. Fig. \ref{fig:one_arm_profile} shows the composite profilometer scan of the structure. 
As before, moving outwards on the inside of the spiral (i.e. turning clockwise) the ``floor'' ascends towards the unetched Ge surface, 
the etching depth (aside from the central well) being of order $1 \, \mu m$. Referring to Fig. \ref{fig:stills}, we also see the ``delamination zone'' 
as the region of bare Ge outside the spiral line in Fig. \ref{fig:one_arm_pic}, with a depth of engraving of order $100 \, nm$ 
(Fig. \ref{fig:one_arm_profile}). \\ 
For quantitative analysis we turn to the line scans from which Fig. \ref{fig:one_arm_profile} is constructed. 

\begin{figure}[H]
	\centering
	\includegraphics[width=3.5 in]{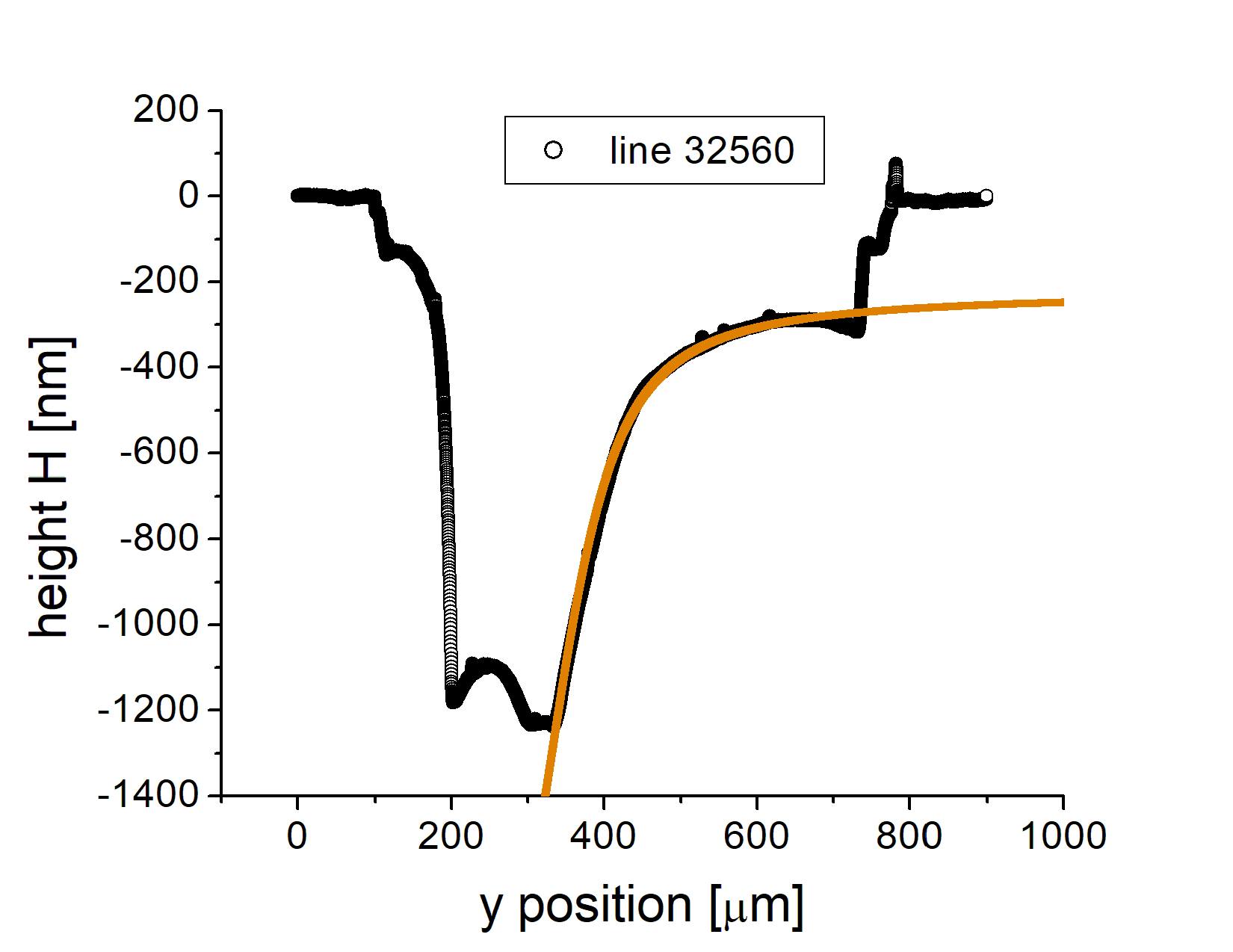}
	\caption{Profilometer scan (H vs y) corresponding to the line $x = 32560$ in Fig. \ref{fig:one_arm_profile}. Note that 
	the y coordinate is pointing down in Fig. \ref{fig:one_arm_profile}. Beyond the central pit ($200 < y < 336 \, [\mu m]$), 
	we focus on the profile for ($336 < y < 720$). This segment corresponds to the region swept by the Ge-metal contact line 
	as the spiral grows (see Figs. \ref{fig:one_arm} and \ref{fig:stills}). The orange line corresponds to the model (\ref{eq:y_H_model}), 
	described in the text. Here it is plotted with parameter values: $c = 100$ , $\alpha = 230$ , $y_0 = 340$ , $H_0 = - 220$, 
	and the spiral parameters $\gamma = 114$ , $\theta_0 = - 210 \, deg$. }
	\label{fig:Profile_32560}
\end{figure}

Fig. \ref{fig:Profile_32560} shows one example, corresponding to the line $x = 32560$ in Fig. \ref{fig:one_arm_profile}. What is plotted 
is the height $H$ of the Ge surface vs the $y$ coordinate of Fig. \ref{fig:one_arm_profile}, zero height being the unetched surface. 
For $0 < y < 100 \, [\mu m]$ we see the flat, unetched, metal covered Ge surface (compare with Fig. \ref{fig:one_arm}). 
For $y > 100$ we enter the delamination zone, with an initial steep $\sim 130 \, nm$ dip ($100 < y < 110$), a plateau 
($H = -130 \, [nm] \, , \,  110 < y < 140 \, [\mu m]$), and a deep canyon which brings us to the spiral line at 
$y = 200 \, [\mu m] \, , \, H = - 1180 \, [nm]$. For $200 < y < 336 \, [\mu m]$ we traverse the bottom of the well 
which corresponds to the circular central structure visible in Fig. \ref{fig:one_arm_pic}. \\ 
Below we focus on the region $336 < y < 720$, which we describe quantitatively. 
This is the etched region created by the metal-Ge contact line sweeping across the surface following the path of the crack in the metal film 
which defines the spiral line (see Fig. \ref{fig:stills}). Beyond the spiral line ($y > 720$) we see again the delamination zone 
($720 < y < 780$), with a similar structure as before, including the small plateau ($740 < y < 760 \, [\mu m] \, , \, H = - 114 \, [nm]$), 
and finally the flat unetched surface ($y > 780$). For our purposes, and referring to Fig. \ref{fig:stills}, we may divide the etch pattern 
into 3 distinct regions: the region of the etch pit / central well (I)
(square structure in Fig. \ref{fig:stills}, round central structure in Fig. \ref{fig:one_arm_pic}); the outer delamination region (III), 
and region II, between the central well and the (spiral) path of the crack in the metal film. The mechanisms involved in the etching 
process are clearly different for the three regions; here we focus on region II, corresponding to ($336 < y < 720$) in 
Fig. \ref{fig:Profile_32560}, as mentioned before. Referring to Fig. \ref{fig:stills}, we assume that etching happens at the miving 
metal-Ge contact line, and that the amount of material removed and thus the local depth of etching is proportional to the local 
``residence time'' of the metal-Ge contact line and therefore is inversely proportional to the local velocity of the contact line 
perpendicular to itself. As the (straight) contact line sweeps out a sector of the spiral, the local velocity is small close to the center 
well and large further away from the well, and correspondingly the etched Ge surface slopes upwards moving radially away 
from the well. The growth dynamics of the spiral line (the crack in the metal film) was explored in the previous section 
(eq. (\ref{eq:crack_trajectory})); in parametric form: 

\begin{equation}
	R(t) = a \, t + R_0 \quad , \quad \theta (t) = \gamma \, \text {ln} \, t + \theta_0 
	\label{eq:crack_trajectory_2}
\end{equation}

(the constant $R_0$ sets the start of spiral growth at time $t = -R_0 / a$). Here $(R , \theta)$ are polar coordinates for the spiral line, 
referred to the center of the etch pit as the origin. The azimuthal velocity of the metal-Ge contact line at a distance $r$ from the origin 
is therefore $v_{\theta} (r , t) = r \, d \theta / dt = r \gamma / t$. Our assumption is that at the position $r$ at time $t$ 
(which defines the position $(r , \theta)$) the height of the Ge surface is 

\begin{equation}
	H = - \frac{k}{v_{\theta}} = - \frac{k}{\gamma} \, \frac{t}{r}
	\label{eq:height}
\end{equation}

\noindent where $k$ is a constant related to the rate of etching at the metal-Ge contact line. Referring to Fig. \ref{fig:one_arm_profile}, 
we count $\theta$ clockwise from the y-axis; the position $(x , y)$ on the plane is then: 

\begin{equation}
	\begin{cases} 
	x = r \, sin \, \theta = r \, sin (\gamma \, \text {ln} \, t + \theta_0) \\
	y = r \, cos \, \theta = r \, cos (\gamma \, \text {ln} \, t + \theta_0) 
	\end{cases}
	\label{eq:xy_position}
\end{equation} 

The profilometer cuts (Fig. \ref{fig:one_arm_profile}) are the lines $x = const.$, so that for a given $t$ we have 
$r = c / sin (\gamma \, \text {ln} \, t + \theta_0)$ and thus: 

\begin{equation}
	y = \frac{c}{tan (\gamma \, \text {ln} \, t + \theta_0)} 
	\label{eq:model_y}
\end{equation}

\begin{equation}
	H = - \frac{k}{\gamma c} t \, sin (\gamma \, \text {ln} \, t + \theta_0)
	\label{eq:model_H}
\end{equation}

\noindent In these formulas, $c$ is the $x$ position of the profilometer line (referred to the origin at the center of the spiral), 
$\gamma$ and $\theta_0$ are constants defining the spiral (eq. (\ref{eq:crack_trajectory_2})), and the scale of time (the parameter $t$) 
is also defined by the spiral line through the first equation in (\ref{eq:crack_trajectory_2}). The orange symbols in Fig. \ref{fig:Profile_32560} 
are a plot of eqs. (\ref{eq:model_y}), (\ref{eq:model_H}) for varying $t$, with two modifications. To account for the location of the origin 
of the coordinate system (which in the model (\ref{eq:model_y}), (\ref{eq:model_H}) is at the center of the etch pit whereas in the 
plots of Fig. \ref{fig:one_arm_profile} and Fig. \ref{fig:Profile_32560} the etch pit is at $(x_0 \approx 32660 , y_0 \approx 340)$) we introduce 
in eqs. (\ref{eq:model_y}), (\ref{eq:model_H}) corresponding shifts in the coordinate $y$ ($y \rightarrow (y - y_0)$) and the constant 
$c$ ($c = -(\text{line position} - x_0)$). In addition, the step in surface height along the path of the crack in the metal film 
(i.e. along the spiral line; in Fig. \ref{fig:Profile_32560} it is at ($720 < y < 780$)) is due to processes not included in the model, 
so we introduce a corresponding step in the height of the surface $H \rightarrow (H - H_0)$; 
for the plot of Fig.\ref{fig:Profile_32560} $H_0 = - 220$. Finally the orange symbols in Fig. \ref{fig:Profile_32560} correspond to: 

\begin{equation}
	\begin{cases} 
	y = c / tan (\gamma \, \text {log} \, t + \theta_0) + y_0 \\
	H = - (\alpha / c) \, t \, sin (\gamma \, \text {log} \, t + \theta_0) + H_0
	\end{cases}
	\label{eq:y_H_model}
\end{equation} 

\noindent where we have used base 10 Logarithms and the constant $\alpha$ is then $\alpha = (k / \gamma) \text{ln} 10$. \\ 
For the spiral line (eq. (\ref{eq:crack_trajectory_2})), also using base 10 logs, we find $\gamma = 114$ , $\theta_0 = - 210 \, deg$. 
The orange symbols in Fig.\ref{fig:Profile_32560} are obtained with the parameter values: $c = 100$ , $\alpha = 230$ , $y_0 = 340$ , 
$H_0 = - 220$. We see that the model reproduces the engraving topography quite well {\it in its region of validity} (region II). 
On the side of large $y$ ($y > 720$ in Fig.\ref{fig:Profile_32560}), this region is cut off by the steep engraving corresponding to 
the path of the crack in the metal film; on the side of small $y$ ($y < 336$ in Fig.\ref{fig:Profile_32560}) it is cut off by the central 
annular region visible in Fig. \ref{fig:one_arm_pic}. \\ 
A different profilometer cut of the same spiral should be represented by the same eqs. (\ref{eq:y_H_model}), with an updated value 
for $c$ (the $x$ coordinate of the profilometer line), possibly an adjusted value for $H_0$, and all other parameters the same, 
in particular the constant $\alpha$. 

\begin{figure}[H]
	\centering
	\subfigure[\label{fig:Profile_32500}]{\includegraphics[width=3.5 in]{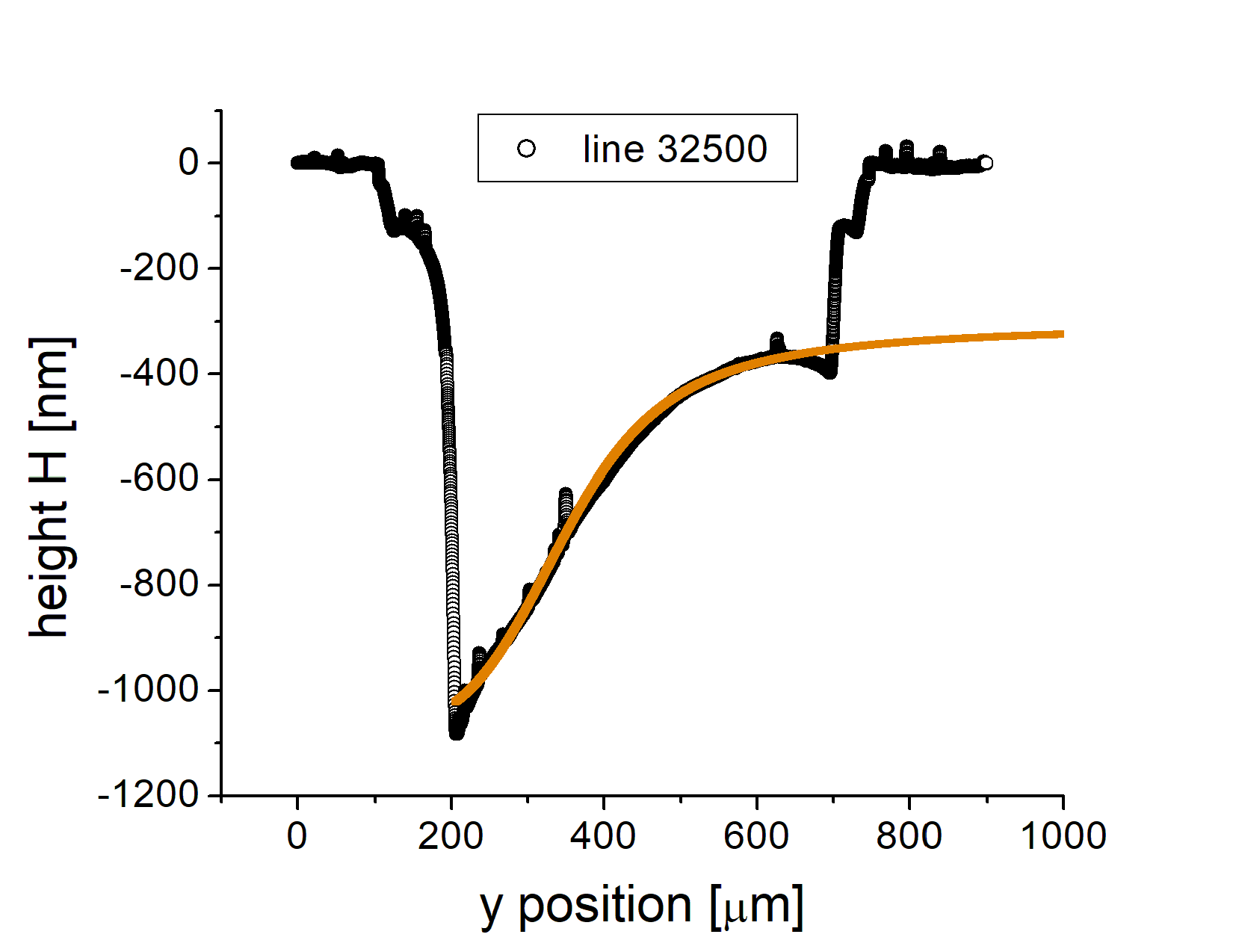}}
	\subfigure[\label{fig:Profile_32440}]{\includegraphics[width=3.5 in]{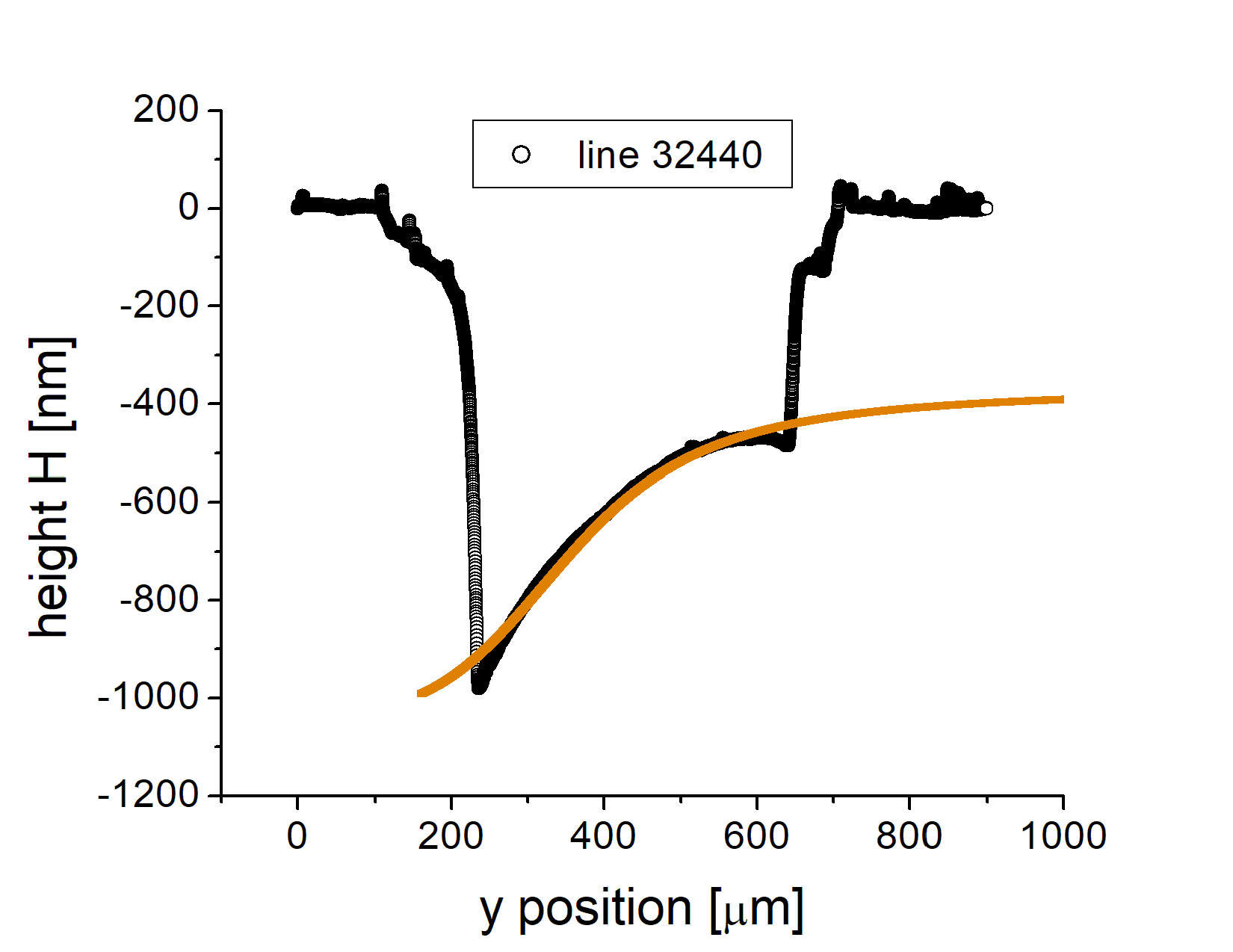}}
	\caption{(a) Profilometer scan corresponding to the line $x = 32500$ in Fig. \ref{fig:one_arm_profile}. 
	The orange line is the model (\ref{eq:y_H_model}) with parameter values: $c = 180$ , $\alpha = 185$ , $H_0 = - 300$ 
	and all other parameters the same as in Fig. \ref{fig:Profile_32560}. \\ 
		(b) Profilometer scan corresponding to the line $x = 32440$ (yellow line in Fig. \ref{fig:one_arm_profile}). 
		The model (\ref{eq:y_H_model}) is plotted with parameter values: $c = 240$ , $\alpha = 210$ , $H_0 = - 360$ 
		and all other parameters the same as in Fig. \ref{fig:Profile_32560}. }
	\label{fig:Profiles_2_3}
\end{figure}

Fig. \ref{fig:Profile_32500} shows a different profilometer cut (the line $x = 32500$ in Fig. \ref{fig:one_arm_profile}), further 
away from the central well. The sloping floor is correspondingly less steep compared to the previous cut.  
The orange symbols represent (\ref{eq:y_H_model}) with $c = 180$ , $\alpha = 185$ , $H_0 = - 300$ and all other parameters the same 
as in Fig. \ref{fig:Profile_32560}. To obtain the rather good agreement, we moved $c$ by 80 units (instead of the nominal 
$60 = 32560 - 32500)$ and slightly adjusted $\alpha$ (from 230 to 185). Fig. \ref{fig:Profile_32440} shows a third profilometer cut, 
60 units to the right of the previous one, with the corresponding plot of eq. (\ref{eq:y_H_model}). Here we moved $c$ by the nominal 
60 units ($32500 - 32440$), and again adjusted $\alpha$ slightly: $c = 240$ , $\alpha = 210$ , $H_0 = - 360$ and all other parameters 
the same. In summary, we can represent all three profilometer cuts (and therefore presumably all lines in between) through 
the model (\ref{eq:y_H_model}), with only relatively small adjustments to the parameter $\alpha$ (at the level of $\pm 10 \%$). 
The agreement gives us confidence that we understand the mechanism for pattern formation in the system; in particular, the one 
physical assumption of the model, namely that the local etching rate is inversely proportional to the local velocity of the 
metal-Ge contact line, seems consistent with the measurements. \\

\section{Materials and Methods}

\noindent {\bf Sample preparation.} The polished p-doped Ge wafer, $127 \, \mu m$ thick, $4 \, inch$ diameter, 
with [100] surface crystal orientation, is $O_2$ plasma cleaned for $10 \, min$. Immediately afterwards the metal layers are 
deposited on the Ge surface by electron beam evaporation:  $4 - 20 \, nm$ of Cr, depending on the sample, and $4 \, nm$ of Au. 
For the logarithmic spiral shown in Fig. \ref{fig:Log_spiral} the metal film thickness was $3 \, nm$ of Cr and $4 \, nm$ of Au. 
For Fig. \ref{fig:Lotus_flower} and the logarithmic spirals of Figs. \ref{fig:stills} and Fig. \ref{fig:one_arm}, 
the metal film thickness was $20 \,nm$ of Cr and $4 \, nm$ of Au. 
For Fig. \ref{fig:Lotus_flower} and the logarithmic spirals of Figs. \ref{fig:stills} and Fig. \ref{fig:one_arm}, 
the metal film thickness was $20 \,nm$ of Cr and $4 \, nm$ of Au. 
The Cr layer was deposited at a rate of $1.2 \, nm / min$, the Au at $3 \, nm / min$; the pressure at the start of the 
process was between $10^{-6}$ and $10^{-7}$ Torr. 
While the Cr layer is substantially uniform, the Au will typically form $\sim 50 \, nm$ size islands \cite{Yilin_2}. The wafer thus 
prepared is stored in air. For the experiments, the Ge-metal wafer is cut into $\sim \, 2 \, cm \times 4 \, cm$ rectangular pieces, 
which are rinsed with Acetone, Ethanol and D.I water sequentially to remove potential surface contamination, 
and blow dried with nitrogen. $500 \, \mu L$ of ``etching solution'' is then transferred on the surface of the chip, forming a drop 
covering part of it. The sample is now incubated at room temperature in open air, allowing the drop to evaporate. 
Within about 8 hours the sample is dry, with salt crystals from the etching solution covering the surface.
The crystallized salt is rinsed off with D.I water and the chip blow dried. A small number of spiral patterns or etch pits may form 
and become visible under the microscope at this step, but the bulk of the patterns are generated during the next incubation step.
In this second step, the chip is re-incubated with another $500\mu L$ of etching solution for $24 - 48$ hours, in a moist chamber 
to prevent evaporation of the drop. During this time the patterns shown in the figures form. After blow drying, we mount 
the chip on a microscope slide for easy handling and storage. All samples were prepared at room temperature. \\ 

\noindent {\bf Etching solution.} We use a mild etching solution, consisting of $1 \, M$ phosphate buffer ($KH_2PO_4$) in DI water, 
$pH 4$, with the addition of $10 \, \mu M$ of a 20 bases single stranded DNA oligomer (Solution 1). The DNA is thiol modified 
at one end, for covalent attachment to the gold. It's presence is ``historical'': the initial patterns were obtained (accidentally) 
with this formulation, so we kept it in subsequent experiments. However, later experiments showed that the presence of DNA 
is not obligatory for forming the patterns. \\ 
For the samples presented in this paper, all patterns were etched in the presence of DNA at $pH\,4$, except for the sample 
shown in Fig. \ref{fig:one_arm}, which was etched in the presence of DNA at $pH\,9$.\\

\noindent {\bf Visualization.} The patterns on the dry samples were observed with a metallurgical microscope 
(i.e. with white light illumination through the objective) at relatively low magnification 
($5 \, \times$, $10 \, \times$, or $20 \, \times$ objective), and recorded with a digital camera attached to the microscope. 
Time-lapse movies of the growing patterns (i.e. with the wet samples) were recorded in the same manner, with time intervals 
of 3 to 10 minutes between frames. In this case, we constructed a wet chamber from the Ge chip and a glass cover slip separated by 
$127\, \mu m$ thick plastic spacers. To prevent evaporation, the chamber was sealed using two-part epoxy. \\ 

\noindent {\bf Strain Measurement} Strain measurements for Fig. \ref{fig:strain} were obtained by analyzing the metal film cones 
in the movie corresponding to Fig. \ref{fig:stills}. The apex of the cone was identified by extrapolating the edge (slant height) of the metal film, and the base was defined at an equal distance down the slant height from the apex. Strain was calculated based on the geometry of the metal film cone according to eq. (\ref{eq:film_strain}) and plotted as a function of distance between the base of the cone 
and the center of the etch pit. \\

\noindent Finally, we note that formation of a specific pattern requires a combination of conditions some of which we control 
(e.g. metal film thickness, etching solution composition) and some which we do not control at present (e.g. presence and location 
of crystal defects). \\

\section  {Discussion} 

\noindent In this report we focus on one of the several different patterns generated by this solid state system, namely the 
Logarithmic spiral. The beauty of these patterns (Fig. \ref{fig:Log_spiral}) reflects their astounding regularity: almost mathematically 
precise, as we have shown (eq. (\ref{eq:crack_trajectory})). This is the more surprising as there are at least four different processes 
which conspire to form the pattern. Referring to Fig. \ref{fig:stills}, there is the central etch pit. Its formation possibly involves 
the translocation of the corresponding metal film patch with the growing depth of the etch pit floor, a process described in MACE 
technology \cite{Rykaczewski2011, Hildreth2012}. Then there is the crack 
at the Ge-metal interface (the metal-Ge contact line), extending in a straight line from the rim of the etch pit to the crack in the metal 
film which defines the spiral line. We have seen that etching along this line proceeds at a rate inversely proportional to the local velocity 
of the line. The crack in the metal film produces a deep groove in the Ge (Figs. \ref{fig:Profile_6400B}, \ref{fig:Profile_75_50}) 
which is the spiral arm. The motion of this crack is driven presumably by a combination of residual (i.e. pre-existing) stress 
in the metal film and the stress created by the etching reaction which removes material from under the metal film. The motion 
of the metal-Ge contact line is presumably similarly driven by a combination of these two effects. Finally, there is the delamination zone, 
where the metal film is still covering the Ge surface but the etching solution has seeped in between the metal and the Ge. The growth 
of the cracks (both the crack in the metal film and the crack which forms the metal-Ge contact line) happens in a regime very different 
from the one often addressed in the physics literature \cite{Fineberg1991, Bouchbinder2010, Svetlizky2014}, 
where the tensile stress is above a critical value and the crack propagates at close to the speed of sound. The driving force for 
the growth of the cracks in the present system is a combination of corrosion and mechanical stress, and the crack velocity is many 
orders of magnitude smaller than the speed of sound in the material, and likely controlled by reaction rates rather than mechanics. 
This mechanism of failure is important in areas of materials science and in geological contexts \cite{Eppes2017}. \\ 
What ties together the different processes which produce the beautiful pattern of Fig. \ref{fig:Log_spiral} ? We believe the answer 
is the presence of an initial singularity, specifically a screw dislocation perpendicular to the Ge surface. It is well established that 
etch pits in Ge and Si single crystals form preferentially at locations where a screw dislocation intersects the sample surface 
\cite{Rhodes1957, Amelinckx_Book}. All our spiral patterns emanate from a central etch pit, which we believe is formed around 
the core of a screw dislocation. Besides causing the formation of the etch pit, the long range strain field due to the dislocation is 
imprinted on the metal film, as we have shown (eq. (\ref{eq:film_strain})). We propose that a stress field of the form 
(\ref{eq:dislocation_strain}), coupled with the etching reaction, guides the trajectories of the cracks and ultimately the formation 
of the spiral. A mathematically precise pattern would then arise from the unique far field generated by a specific singularity. \\ 
We have described quantitatively one of the remarkable patterns formed by this solid state system, namely 
the Logarithmic spiral. It is remarkable that an essentially exact geometric shape arises spontaneously from a non equilibrium 
growth process, which itself follows a mathematically precise dynamics (eq. \ref{eq:crack_trajectory})), which moreover 
extrapolates to a singularity back in time (Figs. \ref{fig:r_vs_t} and \ref{fig:theta}). The genesis of the ordered pattern here 
is different from other well studied nonequilibrium systems, such as the ordered pattern of rolls in Rayleigh-Benard convection 
\cite{Bodenschatz2000}, which arises from the boundary conditions. \\ 
We have described the specific mechanisms involved in producing the spiral in some detail. But what is general about 
this pattern forming process ? This system points to the role of singularities. 
If we ask: where does order, as in Fig. \ref{fig:Log_spiral}, arise from ? We have the following answer. 
Ultimately order comes from the Ge crystal lattice, in a roundabout way, necessary to bridge the 6 orders of magnitude in scale 
between the atomic scale of the lattice and the $\sim 100 \, \mu m$ scale of the pattern. Namely, a defect in the crystal lattice, 
a singularity, produces a geometrically well defined long range field, a stress field in this case, which couples to the nonequilibrium 
growth dynamics to produce an ordered structure. There are other examples of the role of singularities in creating ordered structures, 
notably in fluid dynamics. Thermal plumes are coherent structures which arise spontaneously in thermal convection 
\cite{Zocchi1990}; they are well described in their shape, dynamics and interactions by a flow field created by 
singularities (sources and sinks) \cite{Moses1991, Moses1993, Wang2010}. The ordered structure of a vortex line or a vortex ring 
can similarly be attributed to the far field of a corresponding singularity. Though it involves a breathtaking change of scales, 
it is natural to wonder whether the massive black hole known to reside at the center of most or all spiral galaxies 
\cite{Ghez2008, Davis2014} is the singularity connected to the spiral pattern.  \\ 
For the present system, we still lack a theory to predict, for example, the velocity of the cracks and their 
paths, as well as the different patterns displayed by the system. However we can now argue that such a theory should take 
the local residual stress at the Ge-metal interface as initial condition. Since this is a parameter which we do not control in the 
experiments, this explains the difficulty in establishing experimental control parameters for the different patterns, which we 
mentioned in our first report \cite{Yilin_3}. Nonetheless we now view this system as an interesting paradigm of how 
geometrical order can emerge from singularities in a nonequilibrium process.

\bibliography{Yilin_4}

\end{document}